\documentclass[10pt,journal,compsoc]{IEEEtran}

\usepackage{chngcntr}
\usepackage{tabularray}
\usepackage{tikz}
\newcommand\submittedtext{%
  \footnotesize This work has been submitted to the IEEE for possible publication. Copyright may be transferred without notice, after which this version may no longer be accessible.}

\newcommand\submittednotice{%
\begin{tikzpicture}[remember picture,overlay]
\node[anchor=south,yshift=10pt] at (current page.south) {\fbox{\parbox{\dimexpr0.65\textwidth-\fboxsep-\fboxrule\relax}{\submittedtext}}};
\end{tikzpicture}%
}

\ifCLASSOPTIONcompsoc
  \usepackage[nocompress]{cite}
\else
  \usepackage{cite}
\fi
\ifCLASSINFOpdf
\else
\fi

\usepackage{amsmath}
\usepackage{amssymb}
\usepackage{bm}
\usepackage{xcolor}
\usepackage{graphicx}
\usepackage{makecell}
\usepackage{tabularx}
\usepackage{subcaption}

\usepackage[linesnumbered, ruled, vlined]{algorithm2e}
\SetKwComment{Comment}{$\vartriangleright$ }{}
\SetKwRepeat{Do}{do}{while}%
\SetKwProg{Fn}{Function}{}{end}

\usepackage{soul} 
\usepackage{url}

\usepackage{booktabs}
\usepackage{multirow}
\usepackage{multicol}

\usepackage{pifont}

\newcommand{\remove}[1]{}
\usepackage{tabularx} 

\begin{document}
%
\title{Efficient Probabilistic Workflow Scheduling \\for IaaS Clouds}
%
%
%
%

\author{Gabriele~Russo~Russo, Romolo~Marotta, Flavio~Cordari, Francesco~Quaglia, Valeria~Cardellini, Pierangelo~Di~Sanzo
\IEEEcompsocitemizethanks{\IEEEcompsocthanksitem G. Russo Russo, R. Marotta, F. Quaglia and V. Cardellini are with the Department of Civil and Information Engineering, University of Rome Tor Vergata, Italy. E-mail: russo.russo@ing.uniroma2.it, r.marotta@ing.uniroma2.it, Francesco.Quaglia@uniroma2.it, cardellini@ing.uniroma2.it.
\IEEEcompsocthanksitem F. Cordari is with Sapienza University of Rome. E-mail: flavio.cordari@uniroma1.it.
\IEEEcompsocthanksitem P. Di Sanzo is with Roma Tre University. E-mail: pierangelo.disanzo@uniroma3.it.
}
\thanks{}}

\markboth{}%
{Shell \MakeLowercase{\textit{et al.}}: Bare Demo of IEEEtran.cls for Computer Society Journals}
%



\IEEEtitleabstractindextext{%
\begin{abstract}
The flexibility and the variety of computing resources offered by the cloud  make it particularly attractive for executing user workloads. However, IaaS cloud environments pose non-trivial challenges in the case of workflow scheduling under deadlines and monetary cost constraints. Indeed, given the typical uncertain performance behavior of cloud resources, scheduling algorithms that assume deterministic execution times may fail, thus requiring probabilistic approaches. 
However, existing probabilistic algorithms are computationally expensive, mainly due to the greater complexity of the workflow scheduling problem in its probabilistic form, and they hardily scale with the size of the problem instance.
In this article, we propose EPOSS, a novel workflow scheduling algorithm for IaaS cloud environments based on a probabilistic formulation. Our solution blends together the low execution latency of state-of-the-art scheduling algorithms designed for the case of deterministic execution times and the capability to enforce probabilistic constraints. Designed with computational
efficiency in mind, EPOSS achieves one to two orders lower execution times in comparison with existing probabilistic schedulers.
Furthermore, it ensures good scaling with respect to workflow size and number of heterogeneous virtual machine types offered by the IaaS cloud environment. We evaluated the  benefits of our algorithm via an experimental comparison over a variety of workloads and characteristics of IaaS cloud environments.    
\end{abstract}

}

\maketitle

\submittednotice

\IEEEdisplaynontitleabstractindextext

%
\IEEEpeerreviewmaketitle

\IEEEraisesectionheading{\section{Introduction}\label{sec:introduction}}
\IEEEPARstart{T}he Infrastructure-as-a-Service (IaaS) cloud provisioning model allows to quickly access customized pools of computing resources, which can be chosen from a variety of resource types according to a pay-per-use pricing model. 
However, identifying the resource pool best suited to minimize the execution time or the monetary cost for running a given user workload is not trivial.
It requires to solve an optimization problem, whose complexity is exacerbated by various factors, like the possibility to have pools composed of heterogeneous resources, and the wide variety of resource types and prices currently offered in the IaaS cloud-vendor market. The problem is even more challenging 
for workflows made of multiple tasks. In this case, tasks can be distributed over several computing resources, but the workflow execution schedule must not violate task dependencies (as 
when tasks receive input data produced by other tasks). These kinds of workloads are generally modeled through Directed Acyclic Graphs (DAGs) \cite{scientific_workflow_characterization}, where each node  represents a task, and an edge from a node \textit{a} to node a \textit{b} represents a dependency between task \textit{a} and task \textit{b}. 
DAGs are used in a variety of application domains, in particular for modeling many scientific workloads \cite{schduling_algorithms_taxonomy_2018}.

With the exception of some simple scenarios, the DAG scheduling problem is known to be NP-complete \cite{10.1145/344588.344618}. Consequently, various polynomial-time heuristics have been proposed (e.g. \cite{zhao2020dag, Evolutionary_DAG_multiprocessor_1994, Genetic_DAG_multiprocessor_1994, Incremental_Genetic_DAG_multiprocessor_2004}). Also, it has been investigated in 
different  
computing environments, like multiprocessor servers (e.g. \cite{zhao2020dag, Genetic_DAG_multiprocessor_1994, Incremental_Genetic_DAG_multiprocessor_2004 }), grid environments (e.g. \cite{Genetic_DAG_Grid_Scheduling_2009, Stochastic_DAG_Grid_Scheduling_2011, DAG_Grid_Scheduling_2017}), and the more challenging scenario of IaaS cloud environments (e.g.~\cite{VERMA20171, calzarossa2021, dyna}). In this latter case, the problem consists of finding a configuration of a pool of virtual machines (VMs) and a task assignment plan (that specifies how task executions are assigned to VMs), 
such that a) task dependencies are not violated and b) one or more target metrics are optimized under a set of constraints. Metrics to be optimized and constraints can change based on the specific problem formulation. The most common formulation requires to find a solution that minimizes the monetary cost for executing the workflow, under a time constraint (deadline) on the workflow execution time \cite{dyna}. More in general, the problem can be formulated as a multi-objective constrained optimization problem, which requires to find of a Pareto-optimal set of solutions~\cite{calzarossa2021}, including 
all the solutions such that any of the other candidate solutions has higher monetary cost or higher workflow execution time.

Any approach for solving the DAG scheduling problem in IaaS environments also requires to cope with the shared and virtualized nature of cloud resources. These factors inevitably lead to a non-minimal variability of task execution times~\cite{MEZNI2018139,10.1145/3573009}, thus making them more uncertain, and also affecting the actual monetary cost~\cite{MEZNI2018139, Ristov_7912686}. For example, the studies presented in~\cite{10.14778/1920841.1920902} and~\cite{5708447} measured a variability of CPU performance in AWS EC2 instances while running scientific workflows up to 24\% and 30\%, respectively. The survey in \cite{10.1145/3573009} mentions some studies that report running time variability measurements up to 90\% (e.g. ~\cite{6550282}).

Dealing with uncertain performance behavior requires a probabilistic problem formulation, which allows us to express the desired probability that a given constraint (e.g. a deadline or a maximum monetary cost) is satisfied, in particular under the assumption of arbitrary distributions of execution times. However, as we will show in Section \ref{sec:related}, many proposed solutions for the DAG scheduling problem are based on the assumption of fixed (deterministic) execution times of the workflow tasks. Consequently, they may fail even in the presence of low performance variability. 

In literature there are a  few proposals based on probabilistic models~\cite{calzarossa2021, dyna}, however they suffer from some relevant limitations. First, the proposed algorithms, although based on heuristics, are still computationally expensive. Moreover, they poorly scale when the number of types of VMs offered by the IaaS provider grows, or when the workflow size  increases. We will discuss these aspects with the support of experimental data in Section~\ref{experimental-evaluation}. Finally, not all of them consider
provisioning constraints imposed by IaaS providers, like the limits on the amount of resources of a given type that can be simultaneously used by a user (e.g. Service Quotas in AWS~\cite{aws_service_limits}, Allocation Quotas in Google Cloud~\cite{google_cloud_service_limits}, and Service Limits in Azure~\cite{azure_service_limits}). If these limits are not accounted for while searching the optimal scheduling solution, some VMs may not be available when executing the workflow, and this can lead to violate the deadline.

\remove{Consequently, any solving approach should provide, together with each solution, its probability to be effective. For example,  (e.g. a probability that the solution guarantees that a given deadline is met or that ensures that a given monetary cost will not be exceeded)} 

In this article, we propose \textbf{EPOSS} (Efficient Probabilistic wOrkflow Scheduling for iaas cloudS), a novel scheduling algorithm for IaaS cloud environments based on a probabilistic formulation, which overcomes the above limitations. First, it significantly reduces the latency to find a solution. As we will show, the latency it takes to find a solution is 10 to over 100 times less compared to state-of-the-art scheduling algorithms based on probabilistic formulations. At the same time, it finds solutions as good as, or better than, the other algorithms. Also, it ensures higher scalability and covers a larger set of constraints, which make it particularly suitable to cope with the current provisioning models of IaaS  infrastructures. 

Our work aimed to design an algorithm able to ensure the low execution latency offered by algorithms based on the assumption of deterministic execution times, but that is capable to solve the more complex probabilistic problem. Thus, we designed a solution that leverages the efficiency of the heuristic used by MOHEFT~\cite{moheft}, a well-known scheduling algorithm tailored for deterministic times, combined with a binary search approach and Monte Carlo simulation to identify the best scheduling solutions as a response to probabilistic problem formulations.

In detail, our work offers the following contributions:

\begin{itemize}
\item
EPOSS, an efficient DAG scheduling algorithm for IaaS cloud environments tailored for the probabilistic problem formulation that minimizes the monetary cost under a deadline on the workflow execution time.
\item
P-EPOSS, a parallel version of EPOSS that offers further advantages in terms of algorithm execution time by exploiting multiple processing units.
\item
M-EPOSS, a version of EPOSS capable of solving the multi-objective optimization problem in terms of Pareto optimally, thus minimizing both the workflow execution time and the monetary cost.
\item
An extensive experimental study that compares EPOSS with state-of-the-art algorithms for IaaS environments, considering a variety of scenarios (including different workflows, sets of VM types, task execution time distributions and scalability profiles, and various limits on the amount of resources that users can simultaneously use).
\end{itemize}

The remainder of this article is structured as follows. Section \ref{sec:related} provides an overview of related work. Section \ref{sec:system_model_and_formulation} introduces the system model and the problem formulation. EPOSS, along with its parallel and
multi-objective 
versions are presented in Section \ref{sec:algorithm}. Finally,  Section~\ref{experimental-evaluation} focuses on the experimental results.  

\remove{
DIFFERENZE CON DYNA:
i tempi di esecuzioni di dyna sono elevati per alcune configurazioni
usano solo il limite totale di macchine e non per famimlia
tempi di cpu deterministici
considerano solo 4 tipi di macchine
considerano tempi fissi di cpu (vedi paragrafo inizio 4.1)
}

\section{Related Work}
\label{sec:related}

\begin{table*}[!htbp]
  \centering
  \caption{Relevant characteristics of DAG scheduling algorithms for IaaS cloud environments. A checkmark denotes that a characteristic was fully considered in the algorithm design, ``partially'' denotes it was partially considered and ``extendible'' means that the algorithm was designed to be easily extended to cover the characteristic }
  \label{tab:comparisontable}
  \resizebox{\textwidth}{!}{
    \begin{tabular}{lccccccccccc}
      \hline
      & \multicolumn{2}{c}{\textbf{Optimization metrics}}               & \multicolumn{2}{c}{\textbf{Constraints}}                         
      & \multicolumn{2}{c}{\textbf{Uncertainty}}                     & \multicolumn{2}{c}{\textbf{Resource limits}}
      & \textbf{Formulation}                        \\ \hline
      & \begin{tabular}[x]{@{}c@{}}Workflow\\exec. time\end{tabular} & \begin{tabular}[x]{@{}c@{}}Monetary\\cost\end{tabular} & \begin{tabular}[x]{@{}c@{}}Workflow \\exec. time\end{tabular} & \begin{tabular}[x]{@{}c@{}}Monetary\\cost\end{tabular} & \begin{tabular}[x]{@{}c@{}}Cloud \\resources\\performance \end{tabular} & \begin{tabular}[x]{@{}c@{}}Arbitrary\\ task execution \\ time distributions\end{tabular} & \begin{tabular}[x]{@{}c@{}}Total number\\of VMs or\\vCPUs\end{tabular} & \begin{tabular}[x]{@{}c@{}}Number \\of VMs per\\ VM type\end{tabular}& Probabilistic  \\ \hline
      Abrishami et al.~\cite{ABRISHAMI2013158}                 &                           & \checkmark                & \checkmark                &                           &                           &                           &                            &    &                                          \\ \hline
      Zhu et al.~\cite{ZHU2019880}                 &                           & \checkmark                & \checkmark                &                           &                           &                           &  \checkmark                         & extendible    &                                          \\ \hline
      
      Anwar et al.~\cite{fi10010005}                    &                           & \checkmark                & \checkmark                &                           &                           &                           &                           &    &                                          \\ \hline

      Fard et al.~\cite{10.1145/2996890.2996902}         & \checkmark                & \checkmark                &                           &                           & partially                &                           &                           &    &                                          \\ \hline
      Hu et al.~\cite{HU2018108}                         & \checkmark                & \checkmark                &                           &                           &                           &                           &                           &    &                                          \\ \hline
      Li et al.~\cite{8463469}                           &                           & \checkmark                & \checkmark                &                           & \checkmark                &                           &                           &    &                                          \\ \hline
      Liu et al.~\cite{10.1002/cpe.3942}                 &                           & \checkmark                & \checkmark                &                           &                           &                           &                           &    &                                          \\ \hline
      Meena et al.~\cite{7542128}                        &                           & \checkmark                & \checkmark                &                           & partially                 &                           &                           &    &                                          \\ \hline
      Rodriguez \& Buyya~\cite{10.1109/TCC.2014.2314655} &                           & \checkmark                & \checkmark                &                           & partially                &                           &                           &    &                                          \\ \hline
      Verma \& Kaushal~\cite{VERMA20171} & \checkmark & \checkmark & \checkmark & \checkmark &   &   &   &   &   &   \\ \hline
      Calzarossa et al.~\cite{calzarossa2021}       & \checkmark                & \checkmark                & \checkmark                & \checkmark                & \checkmark                & \checkmark                &                           &                & \checkmark                           \\ \hline
      Zhou et al.~\cite{dyna}                            &                & \checkmark                & \checkmark                &                & \checkmark                & \checkmark                 & partially                 &                & \checkmark                           \\ \hline
      \textbf{EPOSS}                                       & \checkmark                & \checkmark                & \checkmark                & \checkmark                & \checkmark                & \checkmark                & \checkmark                & \checkmark                & \checkmark                \\ \hline
    \end{tabular}
  }
\end{table*}

As already mentioned, 
various heuristics-based polynomial-time resolution algorithms exist for the DAG scheduling problem. These include algorithms designed for different computing environments, like single multiprocessor servers, computer clusters and grids. However, for IaaS  environments, the DAG scheduling problem differs from other cases in that it is characterized by specific metrics to be optimized and constraints~\cite{moheft, VERMA20171}. For example, in IaaS cloud environments, the main metric to minimize is commonly the monetary cost, given the pay-per-use model offered by cloud providers. Along with it, we find the workflow execution time, which is often 
subject to a deadline. The set of constraints and assumptions may vary from case to case (e.g.~\cite{10.1145/2996890.2996902,10.1109/TCC.2014.2314655, VERMA20171, dyna}). Anyway, given the peculiarities of IaaS cloud environments, DAG scheduling algorithms designed for other computing environments are generally inadequate, since they target different objectives, and commonly consider different sets of constraints and assumptions (e.g.  ~\cite{Evolutionary_DAG_multiprocessor_1994, Genetic_DAG_multiprocessor_1994, zhao2020dag, Genetic_DAG_Grid_Scheduling_2009, Stochastic_DAG_Grid_Scheduling_2011}).

Focusing on scheduling solutions designed for IaaS cloud environments, we provide the reader with a comparison between the contribution of our study and other existing scheduling algorithms. We summarise the comparison in Table~\ref{tab:comparisontable}, where we evidenced the relevant characteristics of each algorithm. 
The selected algorithms are collected from various literature articles, and in particular from a recent study~\cite{calzarossa2021}. 
The table shows that the combinations of optimization metrics and constraints considered by the proposed solutions are not always the same. While all algorithms consider the monetary cost as a metric to be optimized, only a subset of them also target the optimization of the workflow execution time. These algorithms define the problem in terms of multi-objective optimization, based on Pareto optimality~\cite{ParetoMultiObjectiveOptimization2005}. Monetary cost and workflow execution time are considered in different way also regarding the problem constraints. 
Most of the proposals assume a constraint (a deadline) on the workflow execution time, while some proposals also assume a constraint (a maximum budget) for the monetary cost. Overall, only a few of the proposed algorithms---i.e., Calzarossa et al. ~\cite{calzarossa2021}, Verma \& Kaushal~\cite{VERMA20171} and our algorithm---target the optimization of both execution time and monetary cost, while also allowing to establish constraints on both of them. 

 Differences among the various algorithms also regard  the assumption on performance uncertainty in IaaS cloud environments. Indeed, some proposals assume deterministic task execution times both in the algorithm design and evaluation. 
Therefore, they offer no evidence that the proposed algorithms can behave correctly under any performance context in IaaS environments. 

Among the proposals that take performance uncertainty into account, only Calzarossa et al.~\cite{calzarossa2021}, Zhou et al.~\cite{dyna}, and our algorithm
assume that task execution times may be arbitrarily distributed and evaluate the proposed algorithms under such an assumption. 
Other algorithms are based on more restrictive assumptions. For example, Fard et al.~\cite{10.1145/2996890.2996902} assume that task execution times are unknown, but within a known interval, and 
evaluate the algorithm using execution times randomly selected within this interval. Meena et al.~\cite{7542128} assume that the resource performance variation is bounded by a known percentage value, thus leading to a bounded variation of the task execution times. In their experimental study, based on previous empirical observations~\cite{10.14778/1920841.1920902}, the percentage variation bound is assumed to be 24\%, and the values were generated according to a normal distribution with mean 12\% and standard deviation 10\%. A similar approach was previously used by Rodriguez \& Buyya~\cite{10.1109/TCC.2014.2314655}. In the experimental study, they assumed a variation of the task execution times between -10\% and +10\%, generated according to a normal distribution. Anwar et al.~\cite{fi10010005} propose a solution that adapts at runtime to variations of task execution times.
In practice, when a task terminates later than expected, subsequent tasks are rescheduled trying to prevent deadline violations. However,  the unpredictability of task execution time is not considered by their algorithm  when building the initial schedule.

Concerning the limits on the amount of resources that a user can simultaneously use, Zhu et al.~\cite{ZHU2019880} consider a constraint on the maximum total number of VMs, claiming that the approach can be extended also for other kinds of resource limits. Our algorithm considers both the maximum total number of VMs (or vCPU) and the maximum number of VMs for each VM type. Among the others, only Zhou et al.~\cite{dyna} consider the presence of resource limits. However, they consider these limits only in the experimental study, where they assume that a new VM could be needed by the scheduler while the maximum amount of VMs admitted for the user has already been reached during the workflow execution. Anyway, they point out that in such a case it is necessary to wait that a VM is released, since their  algorithm does not cope with such a situation.

Finally, 
concerning the problem formulation, the three algorithms in Table~\ref{tab:comparisontable} that consider arbitrarily distributed task execution times (including ours) are also the only ones that consider a probabilistic formulation.
Differently from a deterministic formulation, a probabilistic one allows us to establish the desired probability that a workflow execution deadline is satisfied or that a given monetary cost is not exceeded. As a result, 
we can more reliably capture and cope with performance uncertainty in IaaS cloud environments.
\remove{Consequently, any solving approach should provide, together with each solution, its probability to be effective. For example,  (e.g. a probability that the solution guarantees that a given deadline is met or that ensures that a given monetary cost will not be exceeded)} 
Overall, Table~\ref{tab:comparisontable} shows that our solution considers all the metrics and covers all the requirements as other exiting algorithms. In addition, it also includes further constraints not fully covered by other solutions. 

In terms of resolution approach, our algorithm notably differs from the others. The algorithm by Meena et al.~\cite{7542128} is based on a genetic approach. Rodriguez \& Buyya~\cite{10.1109/TCC.2014.2314655} 
rely on particle swarm optimization, an evolutionary computational technique inspired by the social behavior of bird flocks. Abrishami et al.~\cite{ABRISHAMI2013158} propose a two-phase algorithm, which first determines the deadlines of tasks on the basis of the overall deadline of the workflow, 
and then schedules each task based on its assigned deadline. The solution by Zhu et al.~\cite{ZHU2019880} is partially inspired by multi-resource task packing in distributed systems, an approach based on packing multiple tasks on different servers based on their multiple resource requirements. 

Differently, our algorithm exploits the list-based heuristic of MOHEFT~\cite{moheft} (which will be described in Section~\ref{sec:algorithm}) to preserve high efficiency, coupled with a binary-search based technique and Monte Carlo evaluation to cope with the probabilistic problem formulation. Consequently, our proposal clearly differs from the other two algorithms based on probabilistic formulations. Indeed, the algorithm designed by Calzarossa et al. ~\cite{calzarossa2021} is based on a genetic approach. The only similarity with our algorithm is the use of Monte Carlo simulation. However, as we will show, our solution requires a dramatically smaller number of Monte Carlo evaluations.  This advantage allows our algorithm to largely reduce the execution time, which is often a major obstacle to the usability of scheduling algorithms when the solution space grows. The other probabilistic algorithm, by Zhou et al.~\cite{dyna}, relies on an A*-based searching approach to determine the VM type to execute workflow tasks. The usage of this searching approach is due to the potential advantage of pruning for reducing large search spaces. However, in our experimental study we will show important limitations of this approach. Also, we will show the noticeable advantages of our algorithm in terms of execution time and scalability. 

\section{System Model and Problem Formulation}
\label{sec:system_model_and_formulation}

We consider a workflow composed of a set $V$ of computing tasks. Tasks can be executed in whichever order, unless a task $v_j \in V$ receives input data produced by another task $v_i \in V$. In this case, a dependence exists between the two tasks, thus the execution of $v_j$ can start only after the execution of $v_i$ completes. The whole workflow is described through a directed graph, where nodes represent tasks and directed edges represent dependencies. Assuming that there are no cycles and conditional dependencies between tasks, the graph representing the workflow is a Directed Acyclic Graph (DAG), that we denote as $G=(V,E)$, where $V$ is the set of nodes (or tasks), and $E$ is the set of edges (or dependencies). Hence, an edge is a couple $(v_i, v_j) \in V \times V$, meaning that $v_j$ can start only after that $v_i$ completes. We note that, given a set of nodes $\{u, v_1, v_2, \dots , v_n\} \subseteq V$, if a set of edges $\{(v_1, u), (v_2, u), ..., (v_n, u)\} \subseteq E$ exists, then the execution of task $u$ can start only after the executions of all the tasks $v_1, v_2, \dots, v_n$ terminate.

We assume that the VM pool for executing the workflow can be heterogeneous, i.e., it can be composed of different types of VMs. A VM type denotes a VM configuration in terms of (virtual) computing resources (i.e., type and number of CPUs, amount of memory, etc.). We assume that $\Theta$ is the set of the available VM types. 
\remove{VMs types powered by the same resource type (e.g. the same CPU model) are grouped in the same VM family. Thus, the different performance of VMs types belonging to the same family only depends on the different amount of resources, like the different number of CPUs. We note that this VM classification model fits, or can be straightforwardly adapted, to resource models adopted by major IaaS provider, like AWS, Google Cloud and Microsoft Azure. }

We denote as $M^R$ the maximum amount of resources that can be simultaneously allocated to a user. In practice, $M^R$ can represent the maximum amount of VMs or  virtual CPUs (vCPUs), which are the two parameters commonly used by IaaS providers for limiting the total amount of resources that a user can simultaneously use. Additionally, since IaaS providers often set specific limits on the number of VMs for each VM type that a user can simultaneously use, we denote with $M_j^{type}$, $\forall j \in  \Theta$, the maximum number of allowed VMs of type $j$.

A solution of the scheduling problem is a task assignment plan. This specifies the type of VMs included in the pool and the sequence of tasks that have to be executed on each VM of the pool. Formally, a solution $s$ is a list of $n$ couples $(j_i,L_i)$, with $1 \leq i \leq n $, where:
\begin{itemize}
    \item $j_i \in \Theta$ is the type of the $i$-th VM in the pool
    \item $L_i$ is the list of tasks to be executed on the $i$-th VM in the pool. 
\end{itemize}

We note that $L_1 \cup L_2 \cup ... \cup L_n = V$ and $L_1 \cap L_2 \cap ... \cap L_n = \emptyset$.

The monetary cost of a solution $s$, denoted as $C(s)$, depends on the type and the usage time of each VM in the pool. Assuming that $c^{type}_j$ is the usage cost per unit of time of the VM of type $j$, then $C(s)$ is the sum, for all VMs in the pool, of the products between the VM usage time and $c^{type}_j$. The expected usage time of a VM is given by the sum of the execution times of all tasks that are executed on that VM. We remark that, since VM types differ in terms of configuration of computing resources, the execution time of the same task can be different when executed on different VM types. Also, given the typical uncertain performance behavior of cloud resources, the execution time of a task is subject to variability even when executed on a given VM type. Various task execution models have been considered in literature to calculate the task execution time. For example, in ~\cite{dyna, calzarossa2021}, it is calculated as the sum of task CPU time, I/O time and network communication time. A more detailed model has been used, e.g. in ~\cite{KOVALCHUK20132203}, which includes other factors like queuing times. In any way, since different models may be more or less appropriate depending on specific needs, we abstract from the specific factors affecting the task execution time. Thus, without loss of generality, we model execution time of task $v \in V$ running on a VM of type $j \in \Theta$ as a random variable with arbitrary distribution, that we denote as $t_{v,j}$. We note that this implies that the usage time of each VM in the pool and the monetary cost $C(s)$ are also random variables which depend on the task execution times.   

Given a solution $s$, we denote as $T(s)$ the associated workflow execution time, which corresponds to the makespan of G when the tasks are executed according to the task assignment plan of $s$. Consequently, $T(s)$ is also a random variable, and corresponds to the sum of the execution times of the tasks in the critical path of G.

To simplify the presentation, we introduce our scheduling algorithm focusing on the most common problem formulation, i.e., the one that minimizes the monetary cost under a deadline $d$ on the execution time. We name this formulation as single-objective formulation. We then show how the algorithm can be modified to cope with the multi-objective constrained optimization problem formulation, that requires to find a Pareto optimal solution set.

Based on the foregoing, we formalize the single-objective formulation of the problem as follows:
\begin{align}
&\min_{s \in \Omega} &   &\mathbb{E}[C(s)] \label{eq:obj}\\
&\text{subject to}  &   &P(T(s) \leq d) \geq p_T \label{eq:constraint1}\\
&   &   &N^R(s) \leq M^R  \label{eq:res1}\\
&   &   &N_i^{type}(s) \leq M_i^{type}, \forall i \in  \Theta \label{eq:types1}, \end{align}
\noindent where:
\begin{itemize}
    \item $s$ is a solution of the problem and $\Omega$ is the solution space, i.e., the set of solutions whose task executions do not violate the task dependencies of G; 
    \item $\mathbb{E}[C(s)]$ is the expected monetary cost of $s$;
    \item Equation~\eqref{eq:constraint1} ensures that  the probability to satisfy the deadline $d$ is at least equal to $p_T$;
    \item Equation~\eqref{eq:res1} guarantees that the maximum number of VMs (or vCPUs) $N^R(s)$ simultaneously used by $s$ does not exceed $M^R$;
    \item Equations~\eqref{eq:types1}  constrains the maximum number of VMs $N_i^{type}(s)$, for each type $i \in  \Theta$ and simultaneously used by $s$, to be at most $M_i^{type}$.
\end{itemize}

%

\section{Scheduling Algorithm}
\label{sec:algorithm}
As mentioned in Section \ref{sec:introduction}, our scheduling algorithm EPOSS exploits the heuristic of the MOHEFT~\cite{moheft}\cite{cloudmoheft}, a state-of-the-art algorithm for deterministic execution times. First, we introduce MOHEFT, then we present EPOSS and discuss the role played by MOHEFT.

\subsection{MOHEFT}
\label{sec:moheft}

\begin{algorithm*}
\footnotesize
\caption{MOHEFT: Multi-objective workflow scheduling}
\label{alg:moheft}
\DontPrintSemicolon
\KwData{$G=(V,E)$ \Comment*[r]{DAG}}
\KwData{$D=\{T_{v,j}, \forall (v,j) \in V \times \Theta\}$ \Comment*[r]{Task execution times on each VM type}}
\KwData{$K \in \mathbb{N}^{+}$ \Comment*[r]{Number of solutions to compute}}
\KwResult{$S = \{ s_1, \dots, s_K \} $ \Comment*[r]{Pareto front comprising $K$ solutions}}
$S \gets \{ s_1, \dots, s_K \} $, where $s_i=\emptyset$ \Comment*[r]{Create $K$ empty solutions}
$Rnk \gets$ \textsc{B-Rank}($G$)\;
\For{$u \gets 1,|V|$}{
   $S' \gets \emptyset$\;
   \For{$s_k \in S$}{
        $I \gets s_k \cup \{VM_j\}, \forall j \in \Theta$ \Comment*[r]{Consider VMs already in $s_k$ and a new VM for each type}
        \For{$VM_i \in I$}{
            $s \gets \mathrm{\textsc{AddTask}}(s_k, Rnk_u, VM_i)$ \Comment*[r]{Assing task $Rnk_r$ to $VM_i$ in $s_k$}
                $S' \gets S' \cup \{s\}$\;
        }
    }
    $S \gets \mathrm{\textsc{FirstCrowdingDistance}}(S', K, D$)\;
  }
\end{algorithm*}
MOHEFT stands for \emph{Multi-Objective Heterogenous Earliest Finish Time}~\cite{moheft, cloudmoheft}. It is a polynomial-time heuristic algorithm to schedule workflows on heterogeneous machines. MOHEFT extends the well-known HEFT algorithm~\cite{heft}, a list-based heuristic for optimizing the makespan of workflow-based applications. Compared to HEFT, MOHEFT targets the optimization of multiple objectives by determining a Pareto front. Typically, the objectives include the makespan and the monetary cost.

The MOHEFT pseudocode is reported in Algorithm~\ref{alg:moheft}. Input data include the DAG G, the set D of the (deterministic) task execution times on each VM type, i.e., $\{T_{v,j}\}, \forall (v,j) \in V \times \Theta$, and the number $K$ of solutions to compute along the Pareto front. To determine the front, MOHEFT works as follows. It keeps a set $S$ containing up to $K$ partial scheduling solutions, which includes the optimal ones with respect to each of the objectives, i.e., the makespan and the monetary cost, and the other ones representing trade-off solutions among them. $S$ is initialized with $K$ empty solutions (line~1). Then, all tasks of the workflow are ranked according to the B-Rank metric (see \cite{heft}), which yields a topological 
sorting of the tasks based on their distance to the end of the workflow (i.e., the difference between the expected end time of the workflow execution and the expected task start time). Basically, tasks that must be executed first will have higher rank (line 2). 
MOHEFT schedules one task at a time, starting from the one with highest rank.
It iterates for each task (line~3), and for each iteration it creates a new empty set of candidate solutions $S'$ (line~4).
The set $S'$ is populated by extending the solutions in $S$ with the assignment of the $r$-th task in the ranking. Specifically, for each solution $s_k$, MOHEFT considers a set $I$ of VMs, including all the VMs already in use in $s_k$ and a new VM instance for each type (line~6), generating a new partial solution by adding the $r$-th task to the end of the list of tasks to be executed by each VM in $I$ (lines~7-8).
Then, each generated solution is added to $S'$  (line~9). Before re-starting the cycle for the next task, a new set $S$ is constructed by selecting up to $K$ solutions from $S'$ (line~10), based on their \emph{crowding distance}~\cite{nsgaii}. The crowding distance is a measure of how close a solution is to its neighbor solutions (i.e., how ``crowded'' a certain area of the solution space is). Thus, the $K$ solutions with highest distance are selected. The complexity of MOHEFT is $O(|V|\cdot|\Theta|\cdot K)$.

\subsection{EPOSS Design}
\label{sec:EPOSS}

\begin{algorithm*}
\footnotesize
\caption{Modifications to MOHEFT}
\label{alg:moheft_modified}
\DontPrintSemicolon
\KwData{$d \in \mathbb{R}^+$ \Comment*[r]{Deadline}}
\KwData{$M^R \in \mathbb{R}^+$ \Comment*[r]{Maximum number of VMs or vCPUs}}
\KwData{$\{M_j^{type}, \forall j \in \Theta$\} \Comment*[r]{Maximum number of VMs for each type}}
\KwData{$G=(V,E)$ \Comment*[r]{DAG}}
\KwData{$D=\{T_{v,j}, \forall (v,j) \in V \times \Theta\}$ \Comment*[r]{Task execution times on each VM type}}
\KwData{$K \in \mathbb{N}^{+}$ \Comment*[r]{Number of solutions to compute}}
\KwResult{$S = \{ s_1, \dots, s_K \} $ \Comment*[r]{Pareto front comprising $K$ solutions}}
$S \gets \{ s_1, \dots, s_K \} $, where $s_i=\emptyset$ \Comment*[r]{Create $K$ empty solutions}
$Rnk \gets$ \textsc{B-Rank}($G$)\;
\For{$u \gets 1,|V|$}{
   $S' \gets \emptyset$\;
   \For{$s_k \in S$}{
        $I \gets s_k \cup \{VM_j\}, \forall j \in \Theta$ \Comment*[r]{Consider VMs already in $s_k$ and a new VM for each type}
        \For{$VM_i \in I$}{
            $s \gets \mathrm{\textsc{AddTask}}(s_k, Rnk_u, VM_i)$ \Comment*[r]{Assing task $Rnk_r$ to $VM_i$ in $s_k$}
            \If{$\textsc{Feasible}(s, d, M^R, \{M_j^{type}, \forall j \in \Theta\})$}{
                $S' \gets S' \cup \{s\}$\;
            }
        }
    }
    $S \gets \mathrm{\textsc{FirstCrowdingDistance}}(S', K, D$)\;
  }           
\end{algorithm*}

MOHEFT works under the assumption of deterministic task execution times. In practice, this hardily holds in IaaS  environments. The problem is that in the presence of time variability it is not possible to know in advance the exact time that will be required for each task. To circumvent this problem, one might try to use some kind of approximation, for example the mean (or the median) task execution time. However, any solution which is found based on such data cannot ensure that the deadline will be never violated. More in general, scheduling approaches based on the assumption of deterministic execution times do not allow to control the maximum number of times that a deadline is expected to be violated over a number of workflow executions.

In line with the single-objective formulation of the problem, our algorithm aims at ensuring that the probability to meet the deadline $d$ is at least equal to $p_T$. The basic idea is to exploit MOHEFT using different statistics of the task execution times (i.e., different quantiles) as input for finding a set of candidate solutions, and then identifying the solution  with the minimum cost and that limits the expected number of deadline violations to meet the probability $p_T$. More in detail, EPOSS works by exploring different solutions by means of binary search. For each search step, it calculates the quantile of a given order $\alpha$ ($\alpha$-quantile) of all task execution times and runs MOHEFT using these values as input. Among the solutions returned by MOHEFT, EPOSS selects the one with the lowest monetary cost. Then, this solution is evaluated through a Monte Carlo simulation. This randomly samples task execution times from their expected distributions, simulates a number of workflow executions with the VMs and the task assignment plan determined by the chosen solution, then returns the measured empirical probability of meeting the deadline and the calculated average monetary cost. Based on these results, EPOSS decides whether to continue the search by restricting the quantile search interval to the top half or to the bottom half with respect to $\alpha$. The rationale behind this exploration approach is that higher quantile orders likely lead to more conservative solutions (i.e., with higher probability of meeting the deadline) but with higher monetary cost. Conversely, lower quantile orders likely lead to cheaper solutions but with higher risk of violating the deadline. Once the quantile search interval is smaller than a certain size, the search terminates and the found solution is selected. In what follows, we describe more in detail how EPOSS is implemented and how it works.

\subsubsection{Modifications to MOHEFT}
As a first step, we modified the MOHEFT implementation reported in Algorithm~\ref{alg:moheft} such that a new generated solution $s$ is added to $S'$ only if the workflow execution time does not violate the deadline $d$ and if the amount of resources required by the solution $s$ does not overcome the related limits. In other words, the new solution is added to $S'$ only if it does not violate the  constraints expressed by Equations~\eqref{eq:constraint1},~\eqref{eq:res1} and~\eqref{eq:types1} of the single-objective formulation presented in Section \ref{sec:system_model_and_formulation}.

\begin{algorithm*}
\footnotesize
\caption{Implementation of function Feasible}
\label{alg:Feasible}
\DontPrintSemicolon
\SetKwFunction{FFeasible}{\textsc{Feasible}}
\text{All input data of Algorithm \ref{alg:moheft_modified} plus}
\KwData{$s$ \Comment*[r]{Solution to evaluate}}
\KwResult{whether the given solution is feasible or not}
\SetKwProg{Pn}{Function}{:}{\KwRet}
  \Pn{\FFeasible{$s$, $d$, $M^R$, $M^{type}$  }}{
  \Comment{Functions \textit{EST} and \textit{EFT} return the expected start time and expected finish time, respectively
  }
        $makespan \gets \max_{v \in V}{EFT(v;s)}$\;
        \Comment{Check deadline constraint}        
        \If{$makespan> d$}{
        \KwRet{False}
        }
        \Comment{Check resource-related constraints}
        $Allocations \gets \emptyset$\;
        \For{$VM \in s$}{
        $t_0 \gets \min_{v \in L_{VM}}{EST(v; s)}$
        \Comment*[r]{VM start time}
        $t_1 \gets \max_{v \in L_{VM}}{EFT(v; s)}$
        \Comment*[r]{VM shutdown time}
        $Allocations \gets Allocations \cup \{ (t_0, \theta_{VM}, 1),
        (t_1, \theta_{VM}, -1)\}$ 
        }
        $r \gets 0$ \Comment*[r]{Allocated resources}
        $r_j \gets 0$ \Comment*[r]{Allocated instances of each type $j \in \Theta$}
        $A \gets \mathrm{sort~} Allocations \text{~in ascending time order}$\;
        \For{$(t,j,x) \in A$}{
        $r \gets r + x\*CPU_j$ \;
        $r_j \gets r_j + x$ \;
        \If{$r > M^R$ or $r_j > M_j^{type}$}{
        \KwRet{False}
        }
        }
  }
\end{algorithm*}

The MOHEFT psedudocode with these modifications is reported in Algorithm \ref{alg:moheft_modified}. In the initial part, we added the input data related to deadline and to the two constraints mentioned above. Further, we added an \textit{if} instruction (line~9) that calls the function \textsc{Feasible}($s, d, M^R, M_i$). The latter checks whether the constraints expressed by Equations~\eqref{eq:constraint1},~\eqref{eq:res1} and~\eqref{eq:types1} are satisfied or not. If yes, $s$ is added to $S'$ (line~10). 

The psedudocode of the function \texttt{Feasible}, which checks whether a solution
$s$ satisfies the considered constraints,
is reported in Algorithm \ref{alg:Feasible}. 
The function first checks if the expected workflow makespan
is within the deadline (lines 2--4). For this purpose, the function
computes the makespan as the maximum expected finish time of all
the scheduled tasks. In turn, the expected finish time is simply recursively
computed as described in~\cite{heft}, also accounting for the time spent by each task reading/writing intermediate data.
In this regard, we assume that tasks rely on a shared storage
(e.g. a distributed file system, or an object storage service like AWS S3) to write their output, which will be later retrieved by successor
tasks. To estimate this communication overhead, we divide the expected
output data size  of each task by the estimated network bandwidth of
the machine in use. Such overhead is avoided if the task is co-located
on the same VM with all its successors (i.e., in this case, the local
VM storage can be used to store intermediate data).
If the deadline is met, the function proceeds checking whether the
expected resource usage (i.e., allocated vCPUs and VM instances)  due to workflow execution does not exceed the
maximum amount of resources allowed to the user. To do so, we first build a time-ordered list of VM allocations and de-allocations according to $s$ (lines 6--12), and then
verify that the constraints are not violated at any time (lines 13--17).

\subsubsection{EPOSS}
\begin{algorithm*}
\footnotesize
\caption{EPOSS}
\label{alg:EPOSS}
\DontPrintSemicolon
\KwData{$G=(V,E)$ \Comment*[r]{Workflow}}
\KwData{$\{t_{v,j}, \forall (v,j) \in V \times \Theta\}$, with $t_{v,j}$ random variables \Comment*[r]{Task execution times on each VM type}}
\KwData{$p_T \in [0,1]$ \Comment*[r]{Required probability of meeting the deadline}}
\KwData{$\epsilon > 0$ \Comment*[r]{Search stopping threshold}}

\KwResult{$s_{best}$ \Comment*[r]{Best scheduling solution found}}
$s_{best} \gets \emptyset$\;
$C_{best} \gets \infty$ \Comment*[r]{Cost of best scheduling solution}
$\alpha_l \gets 0$, $\alpha_h \gets 1$ \Comment*[r]{Initial quantile interval}
\Do{$\alpha_h - \alpha_l > \epsilon$}{
    $\alpha \gets \frac{1}{2}\*(\alpha_l+\alpha_h)$\;
    $T_{v,j}^\alpha \gets$ $\alpha$-quantile of $t_{v,j}, \forall (v,j) \in V \times \Theta$ \Comment*[r]{Compute $\alpha$-quantile of task execution times}
    $\bar{S} \gets$ \textsc{MOHEFT}($G$, $\{T_{v,j}^\alpha, \forall (v,j) \in V \times \Theta\}$, $K$)\;
    $\bar{s} \gets  \mathrm{argmin}_{s \in \bar{S}} {  C(s) }$ \Comment*[r]{Least cost MOHEFT solution}
    $\bar{p}_T,\bar{C} \gets \textsc{Evaluate}(G,\bar{s})$\;
    \If{$\bar{p}_T \geq p_T$}{
        $\alpha_h \gets \alpha$\;
       \If{$\bar{C} < C_{best}$}{
            $s_{best} \gets \bar{s}$ \;
            $C_{best} \gets \bar{C}$ \;
       }
    }\lElse{
       $\alpha_l \gets \alpha$
    }
}
\end{algorithm*}

The pseudocode of EPOSS is reported in Algorithm~\ref{alg:EPOSS}. The algorithm keeps track of the best identified solution and its cost, which are initialized as an empty solution with infinite cost (lines~1-2). The quantile search interval is identified through its lowest and highest values, which are initially set as $\alpha_l=0$ and $\alpha_h=1$, respectively (line~3). Then the search starts (line~4), and continues while the search interval is larger than a threshold $\epsilon$, which is included  in the interval $[0,0.5]$  (line~16). For each search step, the algorithm calculates the middle point $\alpha$ of the interval (line~5), then computes the $\alpha$-quantile of the execution time of each task (lines~6) and uses these data as input to our modified version of MOHEFT (lines~7). Among the $K$ returned solutions, it selects the least-cost one $\bar{s}$ (line~8) and calls the function \textit{Evaluate} (line~9). It runs the Monte Carlo simulation to evaluate the solution $\bar{s}$, and returns the related probability $\bar{p}_T$ of meeting the deadline and its average monetary cost $\bar{C}$. Then, the algorithm checks if $\bar{p}_T$ is greater than or equal to $p_T$ (line~10). If yes, it restricts the quantile search interval to the bottom half $[\alpha_l, \alpha]$ (line~11) and, if the estimated monetary cost of the solution $\bar{s}$ is lower than the previously best found solution, then $\bar{s}$ is promoted to the best solution (lines~12-14). Otherwise, it restricts the quantile search interval to the top half $[\alpha, \alpha_h]$ (line~15).

The number of search steps executed by EPOSS depends on the selected value of $\epsilon$ (see line~16). Specifically, it performs $\log_2{\frac{1}{\epsilon}}$ steps. In our experimental evaluation study, we observed that a value of $\epsilon$ equal to $0.02$ is sufficient for achieving with EPOSS an accuracy level as the one reached by the other two state-of-the-art probabilistic algorithms we considered. 
With such a configuration, EPOSS performs 6 search steps (thus it runs MOHEFT at most 6 times), therefore the order of complexity of EPOSS is the same  as MOHEFT, since this is run a constant number of times.

\subsection{Parallel EPOSS}
\label{sec:P-EPOSS}
The EPOSS implementation presented above performs the search steps in sequence. To take advantage of hardware parallelism, we devised a parallel version of EPOSS, that we call \textbf{P-EPOSS}. It does not use the sequential search. Rather, assuming to have $P \geq 2 $ processing units,
P-EPOSS splits the full quantile order interval $[0,1]$ in $P$ uniformly spaced sub intervals, delimited by the sequence of points $\alpha_0, \alpha_1, \alpha_2, \dots\, \alpha_{P-1}, \alpha_{P}$, where $\alpha_0=0$ and $\alpha_{P}=1$. Each sub interval $[\alpha_{i-1},\alpha_i]$, with $i \in [1, P]$, is assigned to one out of $P$ parallel threads, which performs the same sequence of operations in lines~5-9 of Algorithm \ref{alg:EPOSS}. In detail, if a thread is assigned the interval $[\alpha_{i-1},\alpha_i]$, it calculates the $\frac{1}{2}(\alpha_{i-1}+\alpha_i)$-quantile of the task execution times and executes the modified version of MOHEFT using these values as input. Then, it selects the least-costing returned solution and evaluates it via Monte Carlo simulation, which calculates the related probability of meeting the deadline and its average monetary cost. Then, among all the solutions selected by the $P$ parallel processing units, P-EPOSS chooses the one with the lower monetary cost. Subsequently, it selects the sub interval the chosen solution belongs to, and splits this sub interval in $P$ uniformly spaced sub intervals, assigning them to the $P$ parallel threads, which run again. This splitting cycle continues while the size of the sub intervals is greater than the threshold $\epsilon$. Once it terminates, the last chosen solution is returned.

We note that, since P-EPOSS in each step of the cycle splits the interval in $P$ sub intervals, and for each interval it selects the quantile in the middle of the interval, after $c$ cycles the size of the sub intervals is $1/(2 \cdot P^c)$. Thus it terminates after $\log_{2P}{\frac{1}{2\epsilon}}$ steps. In each step, each of the $P$ parallel threads executes the same operations as a single search step of EPOSS. Thus, we can assume that a search step in EPOSS has a duration similar to a splitting step of P-EPOSS. However, the number of search steps required by P-EPOSS to reach the same accuracy as EPOSS is reduced by a factor equal to
\begin{equation}
\frac{\log_2{\frac{1}{\epsilon}}}{\log_P{\frac{1}{2\epsilon}}},
\end{equation}
which corresponds to the speed-up offered by P-EPOSS. For example, with the value of $\epsilon$ equal to $0.02$, as we mentioned above, EPOSS requires 6 search steps, while P-EPOSS requires 3 search steps using 4 parallel threads, and 2 search steps using 8 parallel threads. Detailed measurements that show the advantages of P-EPOSS in terms of algorithm execution time are provided in Section \ref{sec:experimental_evaluation}.

\subsection{Multi-objective Constrained Optimization Problem}
\label{sec:alternative_formulations}
The multi-objective constrained optimization problem can be formulated introducing some additional variables, i.e., the maximum allowed monetary cost $c$ and the probability $p_C$ that the monetary cost is less than $c$. Thus, it can be formulated as:
\begin{align}
&\min_{s \in \Omega} &   &(\mathbb{E}[C(s)], \mathbb{E}[T(s)]) \label{eq:obja}\\
&\text{subject to}  &   &P(T(s) \leq d) \geq p_T \label{eq:constraint31}\\
&                   &   &P(C(s) \leq c) \geq p_C \label{eq:constraint32}\\
&                   &   &N^R(s) \leq M^R  \label{eq:res3}\\
&                   &   &N_i^{type}(s) \leq M_i^{type}, \forall i \in  \Theta \label{eq:types3}. 
\end{align}

With this formulation the problem targets the finding of the set of Pareto optimal solutions. More formally, it is a set $S_P$ of solutions such that for each $s \in S_P$ the constraints expressed by Equations~\eqref{eq:constraint31} and~\eqref{eq:constraint32} are met, and there is no other $s' \in S_P$ such that both the inequalities $\mathbb{E}[T(s)]>\mathbb{E}[T(s')]$ and $\mathbb{E}[C(s)]>\mathbb{E}[C(s')]$ hold true.

To solve the problem based on the multi-objective formulation, we slightly modified EPOSS as in Algorithm~\ref{alg:EPOSSF}, which we name \textbf{M-EPOSS}. M-EPOSS builds a set $A$ of $\epsilon$-quantiles in the interval $[0,1]$ (line~1). Then, for each $\alpha \in A$ (line~2), calculates the $\alpha$-quantile of the  task execution times (line~3) and executes the modified version of MOHEFT using these values as input (line 4). For each $\alpha$, the returned solutions are evaluated via Monte Carlo simulations (lines 5-6), returning in this case the probability $\bar{p}_T$ of meeting the deadline $d$, the probability $\bar{p}_C$ that the monetary cost is less than $c$ and the expected makespan and cost (line~6). Then the solutions that satisfy the constraints expressed by Equations~\eqref{eq:constraint31} and~\eqref{eq:constraint32} are added to a set $F$ of Pareto front candidate solutions (lines~7-8). At the end, all the non-dominated solutions in $F$ (according to their expected makespan and monetary cost) are returned. It is worth observing that M-EPOSS can be easily parallelized, letting each of the $P$ parallel threads execute lines 3-8 for a different $\alpha \in A$.

\begin{algorithm}
\footnotesize
\caption{EPOSS for multi-objective constrained optimization problem}
\label{alg:EPOSSF}

\DontPrintSemicolon
\KwData{$G=(V,E)$ \Comment*[r]{Workflow}}
\KwData{$\{t_{v,j}, \forall (v,j) \in V \times \Theta\}$, with $t_{v,j}$ random variables 
\Comment*[r]{Task execution times on each VM type}}
\KwData{$p_T \in [0,1]$ \Comment*[r]{Probability of meeting the deadline}}
\KwData{$p_C \in [0,1]$ \Comment*[r]{Probability that the monetary cost is less than C}}
\KwData{$\epsilon > 0$ \Comment*[r]{Search stopping threshold}}
\KwResult{$F$ \Comment*[r]{Pareto front}}
$A \gets \{ \alpha_k : k \in \mathbb{N}^{+},  \alpha_k=k \cdot \epsilon,
\alpha_k < 1\}$ \Comment*[r]{Set of $\epsilon$-spaced quantiles in the interval $[0,1]$}
\For{$\alpha \in A$} {
    $T_{v,j}^\alpha \gets$ $\alpha$-quantile of $t_{v,j}, \forall (v,j) \in V \times \Theta$ \Comment*[r]{Compute $\alpha$-quantile of each task execution times}
    $\bar{S} \gets$ \textsc{MOHEFT}($G$, $\{T_{v,j}^\alpha, \forall (v,j) \in V \times \Theta\}$, $K$)\;
    \For{$\bar{s} \in S$}{
    $\bar{p}_T, \bar{p}_C, \bar{T},\bar{C} \gets \textsc{Evaluate}(G,\bar{s})$\; 
    \If{$\bar{p}_T \geq p_T $ and $\bar{p}_C \geq p_C$}{
    $F \gets F \cup (\bar{s}, \bar{T},\bar{C})$ \Comment*[r]{Add solution data}          
    }
    }
    }
remove dominated solutions from $F$\;
\end{algorithm}

\section{Experimental Evaluation} 
\label{experimental-evaluation}
We evaluated our scheduling approach via an extensive experimental study, in which we also compared it with other solutions. We considered a variety of workflow configurations, cloud environment settings, constraint settings and execution time distributions. Compared to previous experimental studies on probabilistic workflow scheduling algorithms \cite{dyna,calzarossa2021}, we used larger sets of workflows and VM types. In this section, we present the details of the experimental study and discuss the achieved results. 

\subsection{Experimental Setting}
This section provides the details about the experimental setting of our study. 

\subsubsection{Workflows}
\label{sec:eval:workflows}
\begin{figure}[t]
\centering

\subcaptionbox{Epigenomics}{\includegraphics[height=4.3cm]{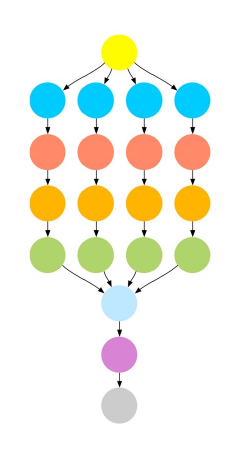}}
\subcaptionbox{Montage}{\includegraphics[height=4.5cm]{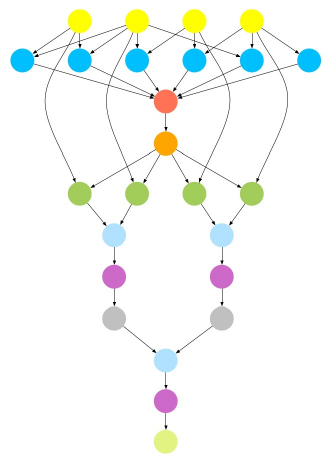}}
\subcaptionbox{SIPHT}{\includegraphics[height=2.9cm]{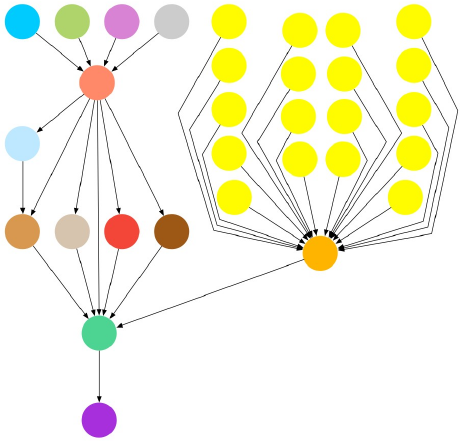}}
\subcaptionbox{LIGO}{\includegraphics[height=3.6cm]{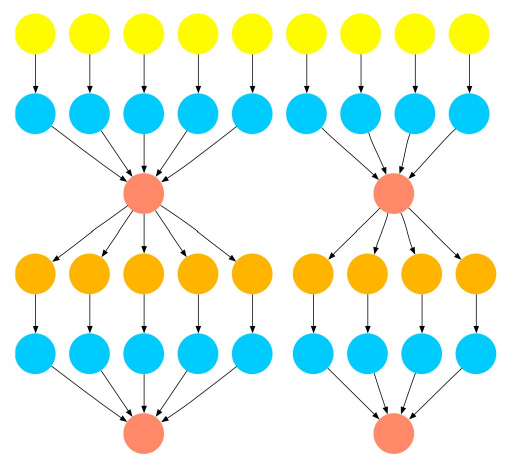}}
\subcaptionbox{CyberShake}{\includegraphics[height=2.2cm]{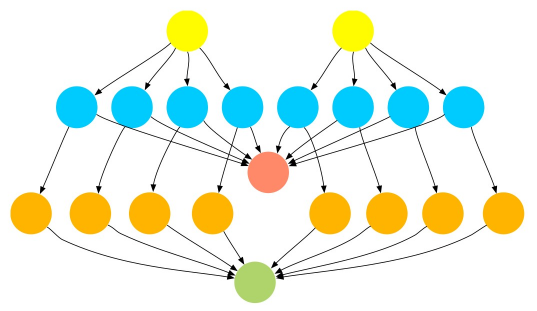}}
\caption{Graphical representations of the workflows used in the experimental study (figures taken from the Pegasus Workflow 
Repository: https://pegasus.isi.edu).
}
\label{fig:workflows}
\end{figure}

The set of workflows we selected includes the five most commonly used ones 
by previous studies on scheduling algorithms \cite{schduling_algorithms_taxonomy_2018}. Specifically, it includes \emph{Epigenomics}, \emph{SIPHT}, \emph{CyberShake}, \emph{Montage} and \emph{LIGO}. These workflows come from different application domains, and together they comprise a variety of common topological structures that characterize scientific workflows. Epigenomics is a workflow used to automate operations in genome sequence processing. SIPHT aims at automating the search for discovering bacterial regulatory RNAs. CyberShake targets the characterization of earthquake hazards. Montage is an astronomy application for creating mosaics of the sky from multiple input images. Finally, LIGO is a workflow designed to analyse gravitational waves.  In Figure~\ref{fig:workflows}, we depict the specific reference structure of each workflow. For a more detailed characterization of the workflows, we refer the reader to~\cite{scientific_workflow_characterization}.

\subsubsection{Task Execution Times}
\label{sec:eval:taskexectimes}
As mentioned, we modeled the task execution time as a random variable. Hence, given a workflow task $v \in V$ and a VM type $j \in \Theta$, we assume that the mean task execution time $\bar{t}_{v,j}$ and the task execution time probability distribution are known. Also based on results presented in previous studies (e.g. \cite{Workflow_real_time_cloud_scheudling_2021, Scheduling_task_cloud_20198, calzarossa2021}), we choose the following probability distributions for our experiments:
\begin{itemize}
    \item \emph{Gamma distribution}, with shape 1 and scale $\bar{t}_{v,j}$  
    \item \emph{Half-Normal distribution}, with mean $\bar{t}_{v,j}$ 
    \item \emph{Uniform distribution}, with support $\left[0,2\cdot\bar{t}_{v,j}\right]$
    \item \emph{Deterministic times}, i.e., always equal to $\bar{t}_{v,j}$      
\end{itemize}
    
Concerning the specific values of the mean task execution times and the estimated size of their outputs, we used the values reported 
 in~\cite{scientific_workflow_characterization}, which provides a detailed characterization of the five workflows we used in our experiments. Data we used are reported in columns \textit{Runtime} and \textit{I/O write} of Tables 3, 4 and 6-8 in the above-mentioned article. For brevity, we do not report here the full set of data. However, to provide an idea about the covered ranges, Table~\ref{tab:meanTaskExecTimes} shows the maximum and the related minimum values of the mean task execution times and of the estimated size of their outputs for each workflow. We remark that, as described in Section~\ref{sec:algorithm}, the mean time to transfer output data from a source VM (where the task that produced data is executed) to a destination VM (where the task that receives data has to be executed) is calculated as the ratio between the total size of data to be transferred and the minimum network bandwidth of the source and destination VMs.

\begin{table}
\centering
\caption{Minimum and maximum values of the mean task execution times and task output sizes as reported in~\cite{scientific_workflow_characterization}}
\label{tab:meanTaskExecTimes}
\resizebox{0.8\columnwidth}{!}{
\begin{tabular}{lrrrr} 
\toprule
\multirow{2}{*}{Workflow}       & \multicolumn{2}{c}{\begin{tabular}[c]{@{}c@{}}Mean task\\execution time (s)\end{tabular}} & \multicolumn{2}{c}{\begin{tabular}[c]{@{}c@{}}Output\\size (MB)\end{tabular}}  \\
                                & \multicolumn{1}{c}{Min} & \multicolumn{1}{c}{Max}                                         & \multicolumn{1}{c}{Min} & \multicolumn{1}{c}{Max}                              \\ 
\cmidrule{2-5}
\multicolumn{1}{r}{Epigenomics} & 0.48                    & 201.89                                                          & 0.90                    & 242.29                                               \\
Montage                         & 0.64                    & 384.49                                                          & 0.10                    & 775.45                                               \\
SIPHT                           & 0.03                    & 3311.12                                                         & 0.03                    & 567.01                                               \\
LIGO                            & 0.13                    & 0.14                                                            & 0.01                    & 0.13                                                 \\
CyberShake                      & 0.55                    & 265.73                                                          & 0.02                    & 176.48                                               \\
\bottomrule
\end{tabular}
}
\end{table}

\subsubsection{VM Types}
\label{sec:eval:vmtypes}

We defined the set of VM types for our experiments based on the characteristics of VM types currently offered by IaaS providers, in particular taking as a reference the VM types offered by AWS EC2\footnote{\url{https://aws.amazon.com/ec2/instance-types/}}. More in detail, we considered VM types belonging to the EC2 families \texttt{c4} (5 instances), \texttt{c5} (8 instances) and \texttt{m5} (8 instances). Similarly to other IaaS providers, in AWS EC2 all VM types of the same family are powered by the same type of hardware (e.g. same CPU model), but the size of resources is different. In particular, each VM type of a family has a different number of vCPUs with respect to the other VM types of the same family. Based on this, we defined a set of 21 different VM types, each one having the relevant characteristics (i.e., number of cores, bandwidth, and price per hour) as one of the above-mentioned EC2 VM types. Data about the set of 21 VM types we considered in our experiments are reported in Table~\ref{tab:VMTypesReferenceSet}. We note that the set includes VM types spanning from 2 to 96 vCPUs, with a range of prices spanning from 0.114 to 5.520 dollars per hour.

\begin{table}
\caption{Reference set of EC2 VM types}
\label{tab:VMTypesReferenceSet}
\centering
\resizebox{0.8\columnwidth}{!}{
\begin{tabular}{clrrr} 
\toprule
Reference & Reference & &Bandwidth & Price\\
EC2 Family & EC2 Type & vCPUs & (Mbps) & (\$/h)\\
\midrule
c4                                                               & c4.large                                                                                     & 2                                                                                    & 62.50                                                                                & 0.114                                                                                     \\
c4                                                               & c4.xlarge                                                                                    & 4                                                                                    & 125.00                                                                               & 0.227                                                                                     \\
c4                                                               & c4.2xlarge                                                                                   & 8                                                                                    & 125.00                                                                               & 0.454                                                                                     \\
c4                                                               & c4.4xlarge                                                                                   & 16                                                                                   & 125.00                                                                               & 0.909                                                                                     \\
c4                                                               & c4.8xlarge                                                                                   & 36                                                                                   & 1250.00                                                                              & 1.817                                                                                     \\
c5                                                               & c5.large                                                                                     & 2                                                                                    & 1250.00                                                                              & 0.097                                                                                     \\
c5                                                               & c5.xlarge                                                                                    & 4                                                                                    & 1250.00                                                                              & 0.194                                                                                     \\
c5                                                               & c5.2xlarge                                                                                   & 8                                                                                    & 1250.00                                                                              & 0.388                                                                                     \\
c5                                                               & c5.4xlarge                                                                                   & 16                                                                                   & 1250.00                                                                              & 0.776                                                                                     \\
c5                                                               & c5.9xlarge                                                                                   & 36                                                                                   & 1250.00                                                                              & 1.746                                                                                     \\
c5                                                               & c5.12xlarge                                                                                  & 48                                                                                   & 1500.00                                                                              & 2.328                                                                                     \\
c5                                                               & c5.18xlarge                                                                                  & 72                                                                                   & 3125.00                                                                              & 3.492                                                                                     \\
c5                                                               & c5.24xlarge                                                                                  & 96                                                                                   & 3125.00                                                                              & 4.656                                                                                     \\
m5                                                               & m5.large                                                                                     & 2                                                                                    & 1250.00                                                                              & 0.115                                                                                     \\
m5                                                               & m5.xlarge                                                                                    & 4                                                                                    & 1250.00                                                                              & 0.230                                                                                     \\
m5                                                               & m5.2xlarge                                                                                   & 8                                                                                    & 1250.00                                                                              & 0.460                                                                                     \\
m5                                                               & m5.4xlarge                                                                                   & 16                                                                                   & 1250.00                                                                              & 0.920                                                                                     \\
m5                                                               & m5.8xlarge                                                                                   & 32                                                                                   & 1250.00                                                                              & 1.840                                                                                     \\
m5                                                               & m5.12xlarge                                                                                  & 48                                                                                   & 1250.00                                                                              & 2.760                                                                                     \\
m5                                                               & m5.16xlarge                                                                                  & 64                                                                                   & 2500.00                                                                              & 3.680                                                                                     \\
m5                                                               & m5.24xlarge                                                                                  & 96                                                                                   & 3125.00                                                                              & 5.520                                                                                     \\
\bottomrule
\end{tabular}
}
\end{table}

\begin{figure}
    \centering
    \includegraphics[width=0.7\columnwidth]{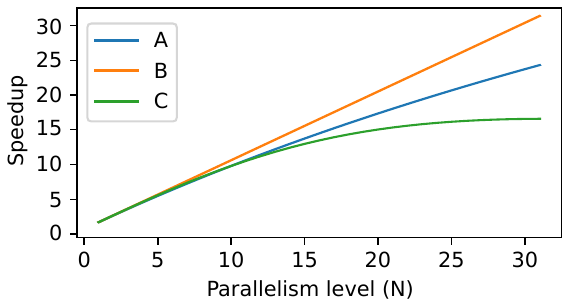}
     \vspace*{-0.3cm} 
    \caption{Task execution times scalability scenarios A, B and C.}
    \label{fig:scalmodels}
\end{figure}

As regards the mean task execution time when the task is executed on different VM types, we used the values reported in~\cite{scientific_workflow_characterization} (see discussion in Section~\ref{sec:eval:taskexectimes})
as baseline for VMs with one vCPU, then we applied the \emph{Universal Scalability Law}~(USL) to estimate how it changes for VM types with a larger number of vCPUs. The USL estimates the relative computation capacity of a machine (or speed-up) with parallelism level $N$ (in our case, corresponding to VM with $N$ vCPUs) as:
\begin{equation}
\label{eq:USL}
C(N) = \frac{N}{1 + \alpha\*(N-1) + \beta\*N\*(N-1)},
\end{equation}
where the coefficients $\alpha$ and $\beta$ depend on the queuing delay to access shared resources of the machine and the delay for ensuring consistency on concurrent shared data accesses, respectively. The combinations of different values of $\alpha$ and $\beta$ correspond to different scalability scenarios. Basically, for the five scientific workloads of our study, we observed that the setting $\alpha=0.01$ and $\beta=0$ (that we identify as scalability scenario \textbf{A}) well fits the shape of their scalability curves. In effect, this is expected with scientific workflows, since the concurrent read/write operations on the same data are typically limited, hence the delay for data consistency is quite irrelevant. However, in our study we considered that different scalability profiles are possible, thus we also included other scalability scenarios. In particular, we considered the antipodal cases with $\alpha=0.01$, $\beta=0$ (scalability scenario \textbf{B}) and with  $\alpha=0.0001$, $\beta=0.001$ (scalability scenario \textbf{C}). The typical shapes of curves for the three scalability scenario are depicted in Figure~\ref{fig:scalmodels}. 

To also account for the different VM families, we added to Equation (\ref{eq:USL}) a specific coefficient, denoted as $\mu_f$. The speed-up with a VM belonging to family $f$ and with $N$ vCPUs is estimated as:
\begin{equation}
C(N, f) = \mu_f \cdot C(N).
\end{equation}

By observing the scalability curves with the three VM families of our study, we found that the best fitting is achieved using $\mu_f=1$ for the \texttt{c5} family, and $\mu_f=0.8$ for the \texttt{c4} and \texttt{m5} families. 

Finally, we remark that we choose to use the USL to estimate the variation of the task execution times rather than using specific values measured on the VM types offered by AWS EC2. Effectively, the specific measured values on the VM types would be strongly dependent on the specific hardware of AWS EC2 VM types. Conversely, our choice allowed our study to span over more variegated scalability scenarios and over larger sets of values.

\subsubsection{Scheduling Algorithms}

For the comparison purposes of our study, in addition to EPOSS, we considered the following algorithms. We selected some baseline algorithms, i.e., HEFT and MOHEFT, that we introduced in Section \ref{sec:related} and in Section \ref{sec:moheft}. Also, we included a variation of HEFT, that we name GreedyCost, which only differs from HEFT in that it minimizes the monetary cost rather than the makespan. We note that although these algorithms do not target probabilistic formulations of the problem (they only consider the mean execution times), they can be relevant for this study to evaluate the advantages offered by probabilistic scheduling approaches. In addition to the above algorithms, we also considered the two state-of-the-art scheduling algorithms which target the probabilistic formulation of the problem, namely Dyna~\cite{dyna} and the algorithm based on the genetic approach presented in~\cite{calzarossa2021} (indicated as Genetic in the following). We described these algorithms in Section~\ref{sec:related}.
We note that the authors of these algorithms decided not to release their source code. For this reason, we implemented the algorithms ourselves based on the details provided by the authors in their articles.

As for the configuration parameters in MOHEFT and EPOSS, we set $K=10$ (as suggested by the authors of MOHEFT in \cite{moheft}) and $\epsilon=0.02$ (see Section \ref{sec:EPOSS}). Regarding the Monte Carlo simulation in EPOSS, we used the statistical stopping criterion based on reaching the 95\% confidence interval for the mean makespan estimation. We note that with all algorithms of our study that use Monte Carlo simulation, we observed that it has a low impact on the total algorithm execution time, generally limited to less than 10\%.

Since all the algorithms of our experimental study can solve the problem based on the single-objective formulation, i.e. the one that minimizes the monetary cost under an workflow execution deadline (see Section~\ref{sec:system_model_and_formulation}), we first focus on the related experimental results. Later, since Genetic was specifically designed to solve the multi-objective constrained optimization problem (see Section~\ref{sec:alternative_formulations}), we present a dedicated comparison study between EPOSS and Genetic.  

\subsubsection{Configurations of Input Parameters}
We run all algorithms changing various configurations of the input parameters. With the  first problem formulation, a single configuration corresponds to a given combination of values of the parameters $\Theta$, $d$, $p_T$, $M^R$ and $M_i^{type}$ $ \forall i \in \Theta$. We varied $\Theta$ over the following different (sub-)sets of the 21 VMs types described in Section~\ref{sec:eval:vmtypes}:

\begin{itemize}
    \item $\Theta_2$ = \{c4.large, c4.xlarge\}
    \item $\Theta_4$ = \{c4.large, c4.xlarge, c4.2xlarge, c4.4xlarge\}
    \item $\Theta_5$ = \{all c4 VM types\}
    \item $\Theta_{13}$ = \{all c4 and m5 VM types\}
    \item $\Theta_{21}$ = \{all c4, c5 and m5 VM types\}
\end{itemize}
where the subscript $x$ of $\Theta_x$ denotes the number of VM types in the set. We varied the deadline $d$ based on the computation demand of each workflow. Specifically, we set $d$ equal to 900 seconds for \emph{CyberShake}, \emph{Epigenomics} and \emph{LIGO}, 2400 seconds for \emph{Montage} and 1800 seconds for \emph{SIPHT}. As for $p_T$, we used three different values, i.e., 0.75, 0.9 and 0.95 (corresponding to the 75, 90, and 95 percentile). Regarding the parameters related to the maximum amount of resources, i.e., $M^R$ and $M_i^{type}$, we remark that Genetic does not consider the presence of such constraints, and Dyna does it only partially (see Table~\ref{tab:comparisontable}). Therefore, for a fair comparison, we first show results obtained by removing these constraints and later we introduce them. 

As for the multi-objective constrained optimization formulation, a configuration also includes the parameters $c$ and $p_C$ (we remark that this formulation requires the monetary cost to be no more than $c$ with probability $p_C$ - see Section \ref{sec:alternative_formulations}).
We set  $c=10$\$ and $p_C=0.9$.

Finally, since EPOSS, Dyna and Genetic are affected by the randomness of the Monte Carlo simulation, we averaged their results over 10 runs for each single configuration, executing each run with a different initial seed of the random number generator.

\subsubsection{Evaluation of Solutions}
\label{sec:evaluation_of_the_solutions}
We evaluated the solutions found by the algorithms by simulating the workflow execution according to the established task assignment plan. More in detail, given a solution $s$ determined by an algorithm for a specific configuration of the input parameters, we executed 10.000 simulation runs, randomly selecting for each run the execution times of each task from the related distribution. Then, we calculated the percentage of runs for which the makespan was below the established deadline $d$. This provided us with an estimation of the probability $P(T(s) \leq d)$ for the solution $s$, that we compared with the desired probability $p_T$ to verify whether or not the solution $s$ satisfies the constraint expressed by Equation (\ref{eq:constraint1}). 
Similarly, with the multi-objective constrained optimization problem, via simulation we also evaluated the probability $P(C(s) \leq c)$ and we compared it with the desired probability $p_C$ to verify whether or not the solution $s$ also satisfies the constraint expressed by Equation (\ref{eq:constraint32}). 
From the simulation results we also calculated the average makespan and the average monetary cost with the solution $s$, that are required to be minimized depending on the different problem formulations. 

\subsubsection{Comparison of Algorithm Execution Times}
To compare the scheduling algorithms in terms of execution times, we executed all algorithms on the same machine, specifically a \texttt{c5.2xlarge} EC2 instance, with 8 vCPUs and 16~GB of memory, with Ubuntu Version 22.04. The reported execution times 
are calculated as average over 10 executions of each algorithm.

\subsection{Results with Single-objective Formulation} 
\label{sec:experimental_evaluation}

The results we present in this section are related to the single-objective formulation of the problem. For Genetic and Dyna, we report the results we achieved with 50,000 internal iterations (as used by their authors in the presented experimental studies) and with 100,000 internal iterations of the algorithm. We denote them as Genetic-50k and Genetic-100k, and as Dyna-50k and Dyna-100k, respectively. For EPOSS, we report the results achieved with both the sequential implementation and the parallel implementation P-EPOSS presented in Section \ref{sec:P-EPOSS}. 

In the first part of this section, we present the aggregated results, calculated by averaging the results we achieved across all the configurations with all values of $p_T$ (0,75, 0.9 and 0.95) and all the five workflows of our study. We focus on the case of task execution times calculated according to the scalability scenario \textbf{A} and distributed according to the Gamma Distribution. Subsequently, we will show the detailed results achieved for the different values of $p_T$ and for the different workflows. Finally, we will focus on the results for the other scalability scenarios and the other distributions of the task execution times. 

\noindent
\subsubsection{Aggregated Results}
\label{sec:experimental_evaluation:aggregated_results}

\begin{figure*}[t]
    \centering
    \includegraphics[width=0.8\textwidth]{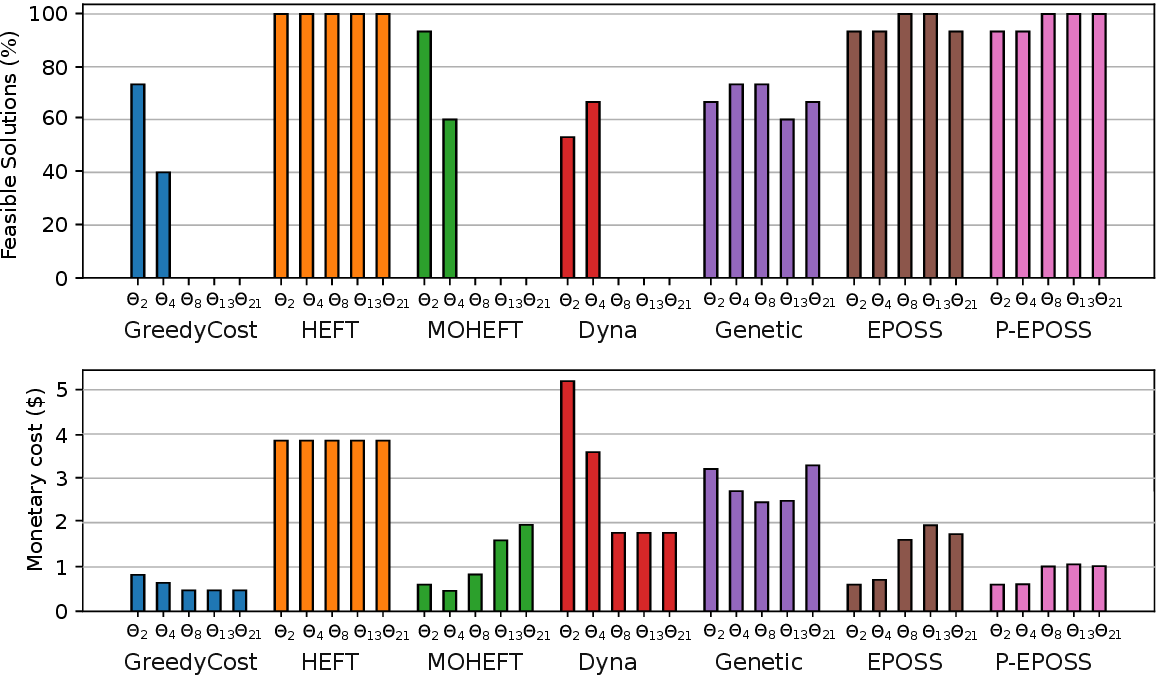}
   \vspace*{-0.1cm} \caption{Aggregated results for different algorithms and different set of VM types}
    \label{fig:aggregatedResults}
\end{figure*}

Figure \ref{fig:aggregatedResults} shows the aggregated results related to the scalability scenario \textbf{A} and with task execution times distributed according the Gamma Distribution. The top graph depicts the percentage of feasible solutions found by the algorithm (i.e., the percentage of configurations for which the algorithm found a solution which ensured the desired probability to meet the deadline), while the bottom graph shows the monetary cost (in dollars) of the found solutions. We analyse data in Figure \ref{fig:aggregatedResults}, together with data related to the average execution times of the different algorithms, which we report in Table \ref{tab:aggregatedExecTime}.

As a first point, we note that the top graph of Figure \ref{fig:aggregatedResults} shows that HEFT always found a feasible solution. We recall that HEFT ignores the monetary cost and only searches for the solution with the lowest makespan. Accordingly, among all the algorithms, we can expect that HEFT likely finds solutions with the highest probability to satisfy the deadline, but with high monetary costs. This is confirmed by the results in the bottom graph of Figure \ref{fig:aggregatedResults}. The opposite behavior characterizes GreedyCost, which ignores the makespan and tries to minimize the monetary cost. Effectively, data show that GreedyCost on average finds the solutions with the lowest cost across all the algorithms. However, only with small sets of VM types it satisfies the deadline a percentage of times higher than 0\%. 

Results achieved by MOHEFT trade off makespan and monetary cost. In fact, they fall in the middle between results of HEFT and of GreedyCost. However, we note that with MOHEFT the percentage of feasible solutions found decreases as the number of VM types increases, dropping to 0\% with 8 or more VM types. These results suggest that, with a non-minimal number of VM types, algorithms like MOHEFT, which relies only on the mean execution times, loose the ability to find feasible solutions for the case of probabilistic formulation of the problem. Indeed, solutions that can guarantee that the mean workflow execution time meets the deadline may not necessarily guarantee that a given percentile of workflow execution times, say 75\% or higher, meet the deadline. Another noteworthy observation is thath the execution times of the three algorithms mentioned so far are generally low compared to the other ones (see Table \ref{tab:aggregatedExecTime}). In the worst case, MOHEFT finished in about 15 seconds. However, the overall results show that these algorithms are generally inadequate to cope with the problem. Indeed, in some cases the solutions they found are associated with high monetary costs, and in the rest of cases only a small percentage of solutions they found are feasible solutions. 

Concerning the other algorithms, a first observation relates the execution time of Dyna. It is not low 
with $Theta_2$, and further grows by 5-6 times moving from $Theta_2$ to $Theta_4$. We note that the study presented by the authors of Dyna in~\cite{dyna} only provides experimental results with a set of 4 VM types. Thus, it does not allow to observe how the algorithm execution time varies depending on larger numbers of VM types. In our study, our implementation of Dyna required more that one hour to terminate with 8 VM types. Also, we noticed that this completion time further increased with more that 8 VM types. Differently, all the other algorithms terminated in about one hour at most, even with the largest size of $\theta$ (i.e., 21 VM types).

We found that the main motivation for the fast growth of the execution time with Dyna is related to its search strategy, which proceeds by ascending the cost of the VM types in the set. In the algorithm of Dyna reported by the authors in~\cite{dyna} at page 5, identified as Algorithm 1, this specifically happens within the \textit{while} cycle starting at line 6, where it searches for a solution that meets the deadline and, if needed, updates the upper bound and the selected solution (lines 12-14).  Within this cycle, when the size of the set of VM types grows, the size of the explored space grows exponentially, even if the algorithm attempts to cut some paths in the search tree. Consequently, in the presence of larger sets of VM types the search time grows fast. Because of the long execution times, in our experimental study the executions of Dyna for all the configurations with 8 or more VM types would have taken an inordinate amount of time.
Therefore, to adopt a fair comparison criterion for all the algorithms of our study, we decided to stop the execution of each algorithm after one hour if it was still running. In this case, we took as result the best configuration found by the algorithm at the time the termination was forced (in Table \ref{tab:aggregatedExecTime}, italicized numbers highlight the cases for which the termination was forced). For this reason, the configuration returned by Dyna with 8 or more VM types often resulted unfeasible under the problem formulation constraints. In particular, with 13 or more VM types, feasible solutions were never found. Thus, for brevity, we omit to report the results of Dyna with 13 or more VM types in the next data tables we present in this article.

\begin{table}
\footnotesize
\centering
\caption{Average algorithm execution times for different algorithms and different set of VM types}
\label{tab:aggregatedExecTime}
\begin{tabular}{lrrrrr}
\toprule
\textbf{Algorithm}      & \multicolumn{5}{c}{\textbf{Execution Time (s)} }   \\
               & $\Theta_2$                & $\Theta_4$ & $\Theta_8$ & $\Theta_{13}$ & $\Theta_{21}$ \\
               \midrule
HEFT           & 0.01                      & 0.01       & 0.02       & 0.04        & 0.08        \\
MOHEFT         & 0.12                      & 0.59       & 2.64       & 7.96        & 15.85       \\
GreedyCost     & 0.01                      & 0.01       & 0.02       & 0.04        & 0.08        \\
Dyna-50k       & 517.95                    & 3550.19    & \textit{3600.58}    & \textit{3600.66}     & \textit{3600.06}     \\
Dyna-100k      & 750.77                    & \textit{3600.22}    & \textit{3600.62}    & \textit{3600.60}     & \textit{3600.16}     \\
Genetic-50k    & 1234.56                   & 1252.84    & 1807.25    & 1791.15     & 1777.75     \\
Genetic-100k   & 2599.34                   & 2647.30    & 3340.96    & 3466.74     & 3457.59     \\
EPOSS         & 2.97                      & 9.62       & 28.26      & 42.12       & 81.86       \\
P-EPOSS & 1.58                      & 5.78       & 16.11      & 28.34       & 61.19       \\
\bottomrule
\end{tabular}
\end{table}

\remove{
To adopt a fair comparison criterion for all the algorithms of our study, we decided to stop the execution of an algorithm if it would not have terminated after one hour. In such a case, we took as result the last best configuration found by the algorithm at the time we forced its termination. For this reason, the configuration returned by Dyna with 8 or more VM types resulted unfeasible under the problem formulation constraints. In particular, with 8 or more VM types feasible solutions were never found. 

We found that the main motivation of the fast growth of the execution time with Dyna is related to its search strategy, which proceeds by ascending cost of the VM types in the set. In the algorithm of Dyna reported by the authors in~\cite{dyna} at page 5, identified as Algorithm 1, this specifically happens within the \textit{while} cycle starting at line 6, where it searches for a solution that meets the deadline and, if needed, updates the upper bound and the selected solution (lines 12-14).
Within this cycle, when the size of the set of VM types grows, the size of the explored space grows exponentially, even if the algorithm attempts to cut some paths in the search tree. Consequently, in the presence of larger sets of VM types the search time grows fast. 
}

Apart from the above aspect related to Dyna, the results show that generally EPOSS found feasible solutions a higher percentage of times compared to Dyna and Genetic for all sizes of the VM type set. Genetic provides better results than Dyna in terms of feasible solutions found. This holds also for the monetary costs, except for 8 or more VM types. However, in these cases the solutions found by Dyna are always unfeasible, and this justifies their lower monetary cost (we remark that Dyna starts searching from the VM types with a lower monetary cost). Another observation is that the improvements achieved by Genetic with 100,000 iterations compared to 50,000 iterations are generally null or small, except for $\theta_{21}$, for which the percentage clearly increases, but also the monetary cost shows a non-minimal increment. Further, the execution time almost doubles in all cases. Therefore, we can conclude that 100,000 iterations offer no obvious advantages over 50,000 iterations. Concerning the monetary costs, with EPOSS they are lower compared to Dyna and Genetic. Moreover, they clearly show the advantages of EPOSS over HEFT. In practice, HEFT always found feasible solutions, but the monetary costs are two to five times higher of the costs of the solutions by EPOSS. 

Data in Table \ref{tab:aggregatedExecTime} show another notable advantage of EPOSS. Indeed, the execution times of Dyna and Genetic with 50,000 iterations are one to as much as three orders of magnitude higher than the execution times of EPOSS (more in detail between 21 and 875 times). This represents a major goal for EPOSS, which proves its greater efficiency compared to the other state-of-the-art probabilistic scheduling algorithms. Finally, data show that the execution times can be further reduced by using the parallel implementation P-EPOSS. We run EPOSS and P-EPOSS on the same machine. However, P-EPOSS has been run with 8 parallel threads, thus exploiting all the 8 vCPUs of the machine, while EPOSS exploited just one vCPU, due to its sequential implementation. The results show that, in terms of feasibility and monetary costs of the solutions, generally P-EPOSS provides results similar to EPOSS. However, it offers a further reduction of the execution times. Such a reduction is not uniform over all the cases, however we measured an average speed-up of 1.47 over all executions. We remark that, as estimated in Section \ref{sec:P-EPOSS}, the maximum theoretical speed up in this case would be $log_28=3$. Given that P-EPOSS still includes some code blocks that cannot be executed in parallel, and considering the presence of shared resources on real hardware, an average speed-up of 1.47 can be expected.

\subsubsection{Detailed Results}
\label{sec:detailed_results}

To provide the reader with a more in-depth view of the experimental results, we report in this section data related to the executions of each single workflow, and separately for the three different values of $p_T$. For space constraints, we focus on the cases with $\Theta_{8}$, $\Theta_{13}$ and $\Theta_{21}$, which are more interesting for the purpose of evaluating the scheduling algorithms of our study. Furthermore, we omit the results obtained with Dyna with more than 8 VM types, because -- as discussed above -- this algorithm fails to find feasible solutions in these settings.
Still for space constraints, in this section we only report data for Epigenomics, while we report data for all the other workflows in the supplemental material of this article.

Table \ref{tab:mainEpigenomics} shows data for Epigenomics with $p_T$ equal to $0.75$, $0.9$ and $0.95$. Each data row in the table refers to the results achieved by an algorithm for a single configuration. In particular, given a solution $s$ returned by the algorithm, the column \textit{Hits} shows the result of the evaluation of the solution $s$ performed as described in Section \ref{sec:evaluation_of_the_solutions}. More in detail, it reports the percentage  of times the makespan was below the desired deadline for the solution $s$, i.e., the percentage of time for which $T(s) \leq d$. If this percentage is less or equal to $p_T$ then the solution $s$ is admissible -- i.e., the constraint $P(T(s) \leq d) < p_T$ of the problem formulation is met. In column \textit{Hits} the values that satisfy this requirement are highlighted in bold. Column \textit{Mon. Cost} shows the average monetary cost (in dollars) of the solutions found. Finally, column {\em Exec. Time (s)} reports the algorithm execution time.

\begin{figure*}[t]
    \centering
    \includegraphics[width=0.8\textwidth]{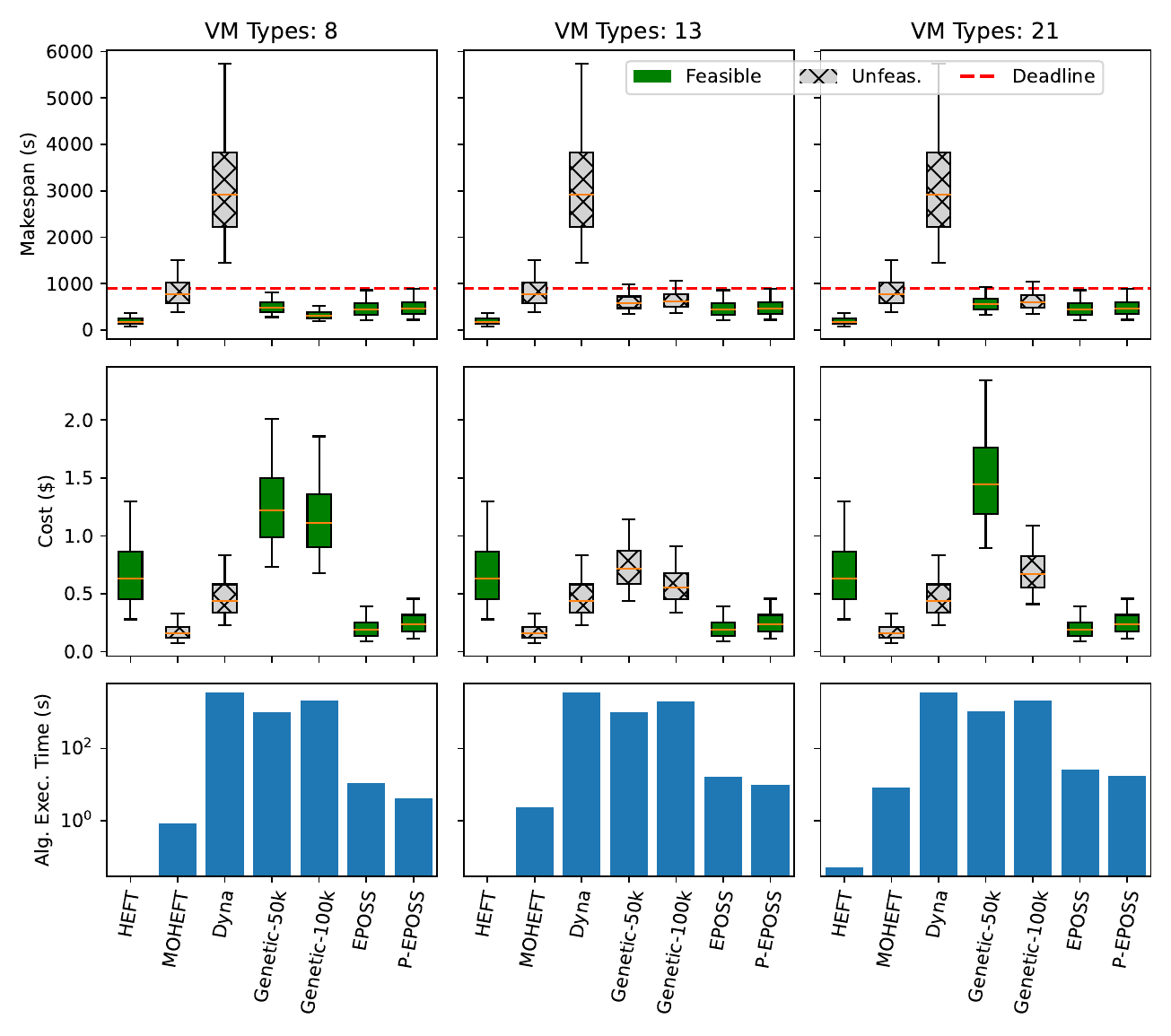}
 \vspace*{-0.3cm} \caption{Graphical representation of the makespan, the monetary cost and the algorithm execution time with the Epigenomics workflow ($p_T=0.9$)}
    \label{fig:mainEpigenomics}
\end{figure*}

For the case with $p_T$ equal to $0.9$ we report additional details in Figure~\ref{fig:mainEpigenomics}, which provides a graphical representation of makespan, monetary cost and algorithm execution time. Specifically, for makespans and costs, the figure uses boxes extending from the first to the third quartile to depict value distribution, along with whiskers extending from
5$^{th}$ to 95$^{th}$ percentile and a horizontal line to indicate the median value. Green boxes refer to scheduling solutions resulting in feasible executions (i.e., for which $P(T(s) \leq d)\geq0.9$), while gray  bars refer to unfeasible solutions. The red dashed line marks the selected deadline (i.e., 900s in the experiment). 

By analysing data in Table \ref{tab:mainEpigenomics} and Figure~\ref{fig:mainEpigenomics}, it is possible to note that overall they reflect trends already observed via the aggregated data reported in Table \ref{tab:aggregatedExecTime}. We found no particular cases that deviated significantly from what we already observed with the aggregate data. This is also confirmed by the data achieved with all the other workflows (available in the supplementary material). Therefore, the behavior of EPOSS observed from the aggregate data is similar with different workflows and different settings. We believe this is a positive result, as it confirms that the benefits of EPOSS are not generally biased toward specific workflows rather than others.  

\begin{table*}[]
\scriptsize
\centering
\caption{Comparison of the scheduling algorithms with the Epigenomics workflow.}
\label{tab:mainEpigenomics}
\resizebox{\textwidth}{!}{
\begin{tabular}{rlrrr|rrr|rrr}

\toprule
 &  & \multicolumn{3}{c}{$p_T=0.75$} & \multicolumn{3}{c}{$p_T=0.9$} & \multicolumn{3}{c}{$p_T=0.95$} \\
VMs & Algorithm  & Hits (\%) & Mon. Cost (\$) &  Exec. Time(s) & Hits (\%) & Mon. Cost (\$) & Exec. Time (s) & Hits (\%) & Mon. Cost (\$) &  Exec. Time (s) \\
\midrule
\multirow[t]{9}{*}{$\Theta_{8}$}
 & HEFT & \textbf{100.0} & 0.69 & 0.02 & \textbf{100.0} & 0.69 & 0.02 & \textbf{100.0} & 0.69 & 0.02 \\
 & MOHEFT & 64.5 & 0.18 & 0.85 & 64.5 & 0.18 & 0.85 & 64.5 & 0.18 & 0.85 \\
 & GreedyCost & 0.1 & 0.13 & 0.02 & 0.1 & 0.13 & 0.02 & 0.1 & 0.13 & 0.02 \\ 
 & Dyna-50k & 0.2 & 0.48 & 3600.35 & 0.2 & 0.48 & 3600.26 & 0.2 & 0.48 & 3600.33 \\
 & Dyna-100k & 0.2 & 0.48 & 3600.41 & 0.2 & 0.48 & 3600.28 & 0.2 & 0.48 & 3600.31 \\
 & Genetic-50k & \textbf{84.4} & 0.78 & 1033.19 & \textbf{94.2} & 1.27 & 1042.62 & \textbf{97.9} & 1.52 & 1020.14 \\
 & Genetic-100k & \textbf{78.0} & 0.55 & 2100.38 & \textbf{100.0} & 1.17 & 2094.36 & \textbf{99.9} & 1.25 & 2064.84 \\
 & EPOSS & \textbf{96.0} & 0.21 & 10.82 & \textbf{96.0} & 0.21 & 10.84 & \textbf{100.0} & 0.24 & 10.33 \\
 & P-EPOSS & \textbf{95.2} & 0.26 & 4.06 & \textbf{95.2} & 0.26 & 4.08 & \textbf{96.6} & 0.29 & 4.15 \\
\midrule
\multirow[t]{7}{*}{$\Theta_{13}$} 
 & HEFT & \textbf{100.0} & 0.69 & 0.03 & \textbf{100.0} & 0.69 & 0.03 & \textbf{100.0} & 0.69 & 0.03 \\
 & MOHEFT & 64.5 & 0.18 & 2.40 & 64.5 & 0.18 & 2.41 & 64.5 & 0.18 & 2.39 \\
 & GreedyCost & 0.1 & 0.13 & 0.03 & 0.1 & 0.13 & 0.03 & 0.1 & 0.13 & 0.03 \\ 
 & Genetic-50k & \textbf{82.4} & 0.75 & 1057.26 & 89.8 & 0.74 & 1046.87 & \textbf{99.5} & 1.23 & 1041.67 \\
 & Genetic-100k & \textbf{97.7} & 1.33 & 2088.53 & 87.3 & 0.58 & 2067.25 & 90.0 & 0.59 & 2108.94 \\
 & EPOSS & \textbf{96.0} & 0.21 & 16.20 & \textbf{96.0} & 0.21 & 16.12 & \textbf{99.2} & 0.35 & 15.51 \\
 & P-EPOSS & \textbf{95.2} & 0.26 & 10.13 & \textbf{95.2} & 0.26 & 10.09 & \textbf{96.6} & 0.29 & 10.21 \\
\midrule
\multirow[t]{7}{*}{$\Theta_{21}$} 
 & HEFT & \textbf{100.0} & 0.69 & 0.05 & \textbf{100.0} & 0.69 & 0.05 & \textbf{100.0} & 0.69 & 0.05 \\
 & MOHEFT & 64.5 & 0.18 & 8.51 & 64.5 & 0.18 & 8.45 & 64.5 & 0.18 & 8.47 \\
 & GreedyCost & 0.1 & 0.13 & 0.05 & 0.1 & 0.13 & 0.05 & 0.1 & 0.13 & 0.05 \\ 
 & Genetic-50k & \textbf{90.8} & 1.48 & 1055.65 & \textbf{90.8} & 1.51 & 1053.54 & 91.5 & 1.10 & 1051.30 \\
 & Genetic-100k & \textbf{77.1} & 0.67 & 2125.96 & 88.7 & 0.70 & 2115.45 & \textbf{96.2} & 0.90 & 2123.56 \\
 & EPOSS & \textbf{96.0} & 0.21 & 26.68 & \textbf{96.0} & 0.21 & 26.64 & \textbf{97.4} & 0.22 & 24.87 \\
 & P-EPOSS & \textbf{95.2} & 0.26 & 17.15 & \textbf{95.2} & 0.26 & 17.24 & \textbf{95.2} & 0.26 & 17.05 \\
\bottomrule
\end{tabular}

}
\end{table*}

\subsubsection{Different Task Execution Time Distributions and Scalability Scenarios}

To verify whether the results provided by the algorithms are affected by different task execution time distributions or by different scalability scenarios, we again analyse some aggregated results. Table~\ref{tab:distributions} shows the results with all distributions other than Gamma, 
i.e., with deterministic times, half-normal distribution and uniform distribution. Specifically, data refer to a configuration with $\theta_8$ and $p_T=0.9$. As expected, with deterministic task execution times all algorithms achieve better results in terms of percentage of times that feasible solutions are found, thanks to the absence of time variability in the execution of all tasks of the workflow. Similar results are achieved with the half-normal distribution. Finally, data with the uniform distribution show that such a case is more challenging for all algorithms, since the percentage drops down for all of them. Entering in the details of each singe algorithm, we note that HEFT, MOHEFT and EPOSS always find a feasible solution with deterministic times and with the half-normal distribution, while Genetic finds 93.33\% of times a feasible solution. However, EPOSS and MOHEFT find solutions with the lower average cost. EPOSS outperforms MOHEFT in the case of uniform distribution. Nevertheless, this distribution represents the harder case also for EPOSS, since the percentage of feasible solutions it finds drops to 73\%, and the average monetary cost increases to 3.52. However, we note that the uniform distribution is representative of an extreme case, in which times are assumed to be randomly distributed within a minimum and a maximum value. This aspect reduces the predictability of task execution times, but in general is rarely observed in real situations (see, e.g., \cite{7019857})) since typically the execution time is not uniformly distributed and is more concentrated around the mean value.

Table~\ref{tab:scalfun} reports the aggregated results for the scalability scenarios A, B and C, introduced in Section~\ref{sec:eval:vmtypes}, using $p=0.9$ and Gamma distribution. Data for Scenario A confirm what we observed above, being an aggregated view of the results we already presented in Section  \ref{sec:experimental_evaluation:aggregated_results}. Data for Scenario B, which represents a simpler scenario where the speed up linearly grows, are quite similar to data for Scenario A. Finally, with the more complex Scenario C, where the speed up decreases beyond a given parallelism level, generally the results are worse with all algorithms. Overall, HEFT shows the highest percentage of feasible solutions across the different scenarios, but also the highest average monetary cost in Scenario A and Scenario B. In these latter cases, the percentage of feasible solutions with EPOSS reaches the value 93.33, and their average monetary costs are lower than the other algorithms. In the more challenging Scenario C, HEFT and EPOSS ensure the highest percentage of feasible solutions. EPOSS also outperforms MOHEFT and Genetic in terms of such a percentage, but the average monetary cost is higher. We believe that this is due to the predictability of the task execution times, which decreases due to the speed up reduction that arises after a certain parallelism level. However, as we already discussed in section Section~\ref{sec:eval:vmtypes}, Scenario C is generally rare in real scenarios, in particular in the case of scientific workflows. 

\begin{table}
\footnotesize
    \centering
    \caption{Result with different task execution time distributions.}
    \label{tab:distributions}
\begin{tabular}{llrr}
\toprule
Distribution    & Algorithm & Feas. (\%) & Mon. Cost (\$) \\
\midrule
\multirow[c]{3}{*}{Deterministic times} & HEFT     & 100.00     & 2.93           \\
  & MOHEFT   & 100.00     & 0.78           \\
  & Genetic  & 93.33      & 1.91           \\
  & EPOSS   & 100.00     & 0.78           \\
\midrule
\multirow[c]{3}{*}{Half-normal}    & HEFT     & 86.67     & 2.89           \\
    & MOHEFT   & 6.67     & 1.03           \\
    & Genetic  & 53.34      & 2.51           \\
    & EPOSS   & 80.00     & 2.22           \\
\midrule
\multirow[c]{3}{*}{Uniform}       & HEFT     & 86.67      & 2.90           \\
       & MOHEFT   & 26.67      & 1.06           \\
       & Genetic  & 46.67      & 3.30           \\
       & EPOSS   & 73.33      & 3.52           \\ 
\bottomrule
\end{tabular}
\end{table}

\begin{table}
\footnotesize
    \centering
    \caption{Results for the different scalability scenarios.}
    \label{tab:scalfun}
    \begin{tabular}{clrr}
\toprule
Scenario & Algorithm  & Feas. (\%) & Mon. Cost (\$) \\
\midrule
\multirow[c]{3}{*}{A} & HEFT    & 100.00 & 2.65 \\
 & MOHEFT  & 80.00 & 0.59 \\
 & Genetic & 80.00 & 1.63 \\
 & EPOSS  & 93.33 & 0.54 \\
\midrule
\multirow[c]{3}{*}{B}& HEFT    & 100.00 & 3.94 \\
 & MOHEFT  & 73.33 & 0.90 \\
 & Genetic & 86.67 & 1.86 \\
 & EPOSS  & 93.33 & 0.78 \\
\midrule
\multirow[c]{3}{*}{C} & HEFT    & 86.67 & 2.16 \\
 & MOHEFT  & 73.33 & 1.13 \\
 & Genetic & 66.67 & 3.64 \\
 & EPOSS & 86.67 & 3.75 \\
\bottomrule
\end{tabular}
\end{table}

\begin{table}
\footnotesize
\centering
\caption{Results with limited provider capacity,    in terms of maximum number of vCPUs that can be allocated concurrently. Missing results indicate that the algorithm could not    determine a feasible schedule with the given constraints.}
\label{tab:capacity}
\begin{tabular}{llrr}
\toprule
$M^R$ & \textbf{Algorithm}  & \multicolumn{2}{c}{\textbf{Mon. Cost (\$)}}                         \\
      &            & $d$=5400s & d=7200s \\
\midrule
25    & Genetic    & –                  & –                  \\
      & HEFT       & –                  & –                  \\
      & MOHEFT     & –                  & –                  \\
      & EPOSS & –                  & 1.16               \\
\midrule
50    & Genetic    & –                  & –                  \\
      & HEFT       & –                  & –                  \\
      & MOHEFT     & –                  & –                  \\
      & EPOSS & 0.93               & 1.13               \\
\midrule
100   & Genetic    & 1.68               & 3.30               \\
      & HEFT       & –                  & –                  \\
      & MOHEFT     & –                  & –                  \\
      & EPOSS & 1.20               & 1.16               \\
\midrule
400   & Genetic    & 1.61               & 2.46               \\
      & HEFT       & 11.86              & 19.45              \\
      & MOHEFT     & –                  & –                  \\
      & EPOSS & 1.31               & 1.16     \\          
\bottomrule
\end{tabular}
\end{table}

\subsubsection{Limits on Resource Amount}
As discussed, experimental data presented above have been achieved by removing from the problem formulation the constraints on the parameters $N^R$ and $N_i^{type}$ for a fair comparison among the algorithms. In this section, we report some results we achieved in the presence of these constraints. We remark that when a resource limit for a given resource type is reached during the execution of a workflow, upon a request of a new resource of the same type it is necessary to wait that one resource of the same type already in use is released. This can lead an increment of the makespan. Since EPOSS can fully accounts for the resource limits, it can potentially find better solutions compared to algorithms like HEFT, MOHEFT and Genetic, which ignore such limits.
In Table~\ref{tab:capacity} we show the average monetary cost of the solution found by the above algorithms with different limit settings. Specifically, we set the maximum number of vCPU that can be simultaneously used $M^R$ equal to 25, 50, 100 and 400, the maximum number of VMs for each type $N_i^{type}$ equal to 10 for each $i \in \theta$ and the deadline $d$ equal to 5400 seconds and 7200 seconds. Data refer to Epigenomics with $\theta_8$ and Gamma-distributed task execution times. The presence of a dash in Table~\ref{tab:capacity} means that the solution returned by the algorithm was unfeasible since it did not satisfy the constraint $P(T(s) \leq d) < p_T$. With the deadline equal to 5400 seconds, no algorithm was able to find a feasible solution for $M^R=25$. Interestingly, with $M^R=50$ only EPOSS found feasible solutions. Genetic found a feasible solutions only with $M^R=100$ and $M^R=400$, but the costs are always higher compared to EPOSS. With the highest deadline, EPOSS found feasible solutions even for $M^R=25$. Conversely, the other algorithms only found feasible solutions for $M^R=100$ and $M^R=400$. In conclusions, the data clearly show the advantages that in some cases EPOSS can provide by accounting for resource limits.

\subsection{Results with Multi-objective Formulation}

\begin{figure}[t]
    \centering
    \includegraphics[width=0.8\columnwidth]{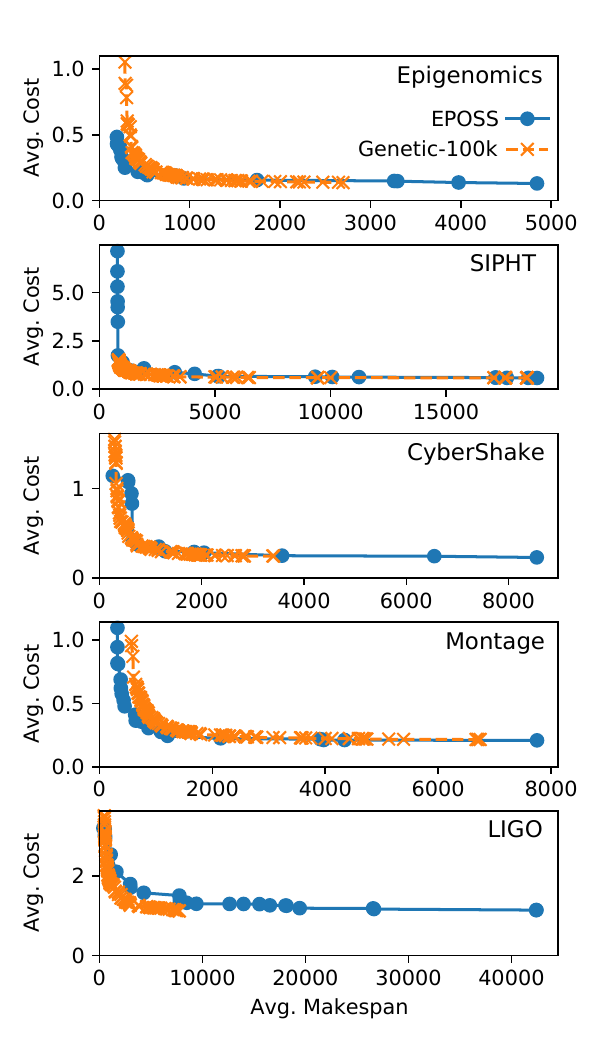}
     \vspace*{-0.3cm}\caption{Pareto fronts calculated by EPOSS and Genetic.}
    \label{fig:frontiers}
\end{figure}

Among the literature algorithms used in our experimental study, the only one that can solve the multi-objective problem and that is based on a probabilistic formulation is Genetic. In this section, we show the results for the multi-objective problem formulation by comparing M-EPOSS (presented in Section \ref{sec:alternative_formulations}) and Genetic. We remark that such a formulation targets the minimization of  both the makespan and the monetary cost, thus it requires to find a Pareto-optimal set of solutions. A graphical comparison of the results is illustrated in Figure~\ref{fig:frontiers}, where the Pareto frontiers returned by M-EPOSS and Genetic are shown for all workflows of our study, for the case with $\theta_8$ and Gamma distribution. 
Results we observed for the other cases are similar. We note that we show results provided by Genetic-100k, since they were slightly better than results by Genetic-50k.

\begin{figure}[t]
    \tiny
    \centering
    \includegraphics[width=0.8\columnwidth]{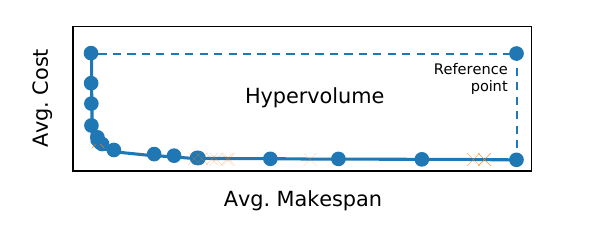}
 \vspace*{-0.3cm} \caption{Example of hypervolume for a Pareto front.}
    \label{fig:hypervolume_example}
\end{figure}

To provide the reader with a clearer quantitative comparison, in Table~\ref{tab:frontier} we also provide data about the associated hypervolume, a reference indicator for evaluating the performance of multi-objective optimization algorithms~\cite{Zitzler2008}. It corresponds to the size of the dominated solution space by the frontier (thus the higher the better), as shown by the example in Figure \ref{fig:hypervolume_example}. The chosen reference point corresponds to the point in the space with the maximum cost and the maximum makespan among all points of the frontier.

\begin{table}[t]
\footnotesize
    \centering
\caption{Relative hypervolume of the frontiers computed by different algorithms
and related execution times.}
\label{tab:frontier}
\begin{tabular}{llrr}
\toprule
Job & Algorithm         &    Hypervol. (\%)     & Alg. Exec. Time (s)  \\
\midrule
Epigenomics & M-EPOSS       & \textbf{100.00} & 17.3   \\
            & Genetic-50  & 97.45  & 351.3  \\
            & Genetic-100 & 98.43  & 1015.8 \\
\midrule
SIPHT       & M-EPOSS       & 100.00 & 73.7   \\
            & Genetic-50k  & 100.95 & 871.7  \\
            & Genetic-100k & \textbf{101.16} & 2473.4 \\
\midrule
CyberShake  & M-EPOSS       & 100.00 & 114.4  \\
            & Genetic-50k  & 100.84 & 570.6  \\
            & Genetic-100k & \textbf{101.37} & 1480.6 \\
\midrule
Montage     & M-EPOSS       & \textbf{100.00} & 110.0  \\
            & Genetic-50k  & 95.55  & 495.3  \\
            & Genetic-100k & 96.34  & 1264.1 \\
\midrule
LIGO        & M-EPOSS       & 100.00 & 231.9  \\
            & Genetic-50k  & 105.06 & 795.3  \\
            & Genetic-100k & \textbf{105.64} & 1946.4\\
    \bottomrule
\end{tabular}
\end{table}

Table~\ref{tab:frontier} shows a comparison in percentage of the calculated hypervolumes for  EPOSS,  Genetic-50k and Genetic-100, and of the algorithm execution times. Overall, we observe that the front hypervolume for EPOSS and Genetic are very close, with a difference never higher than 5\%. Among the five considered workflows, neither of the two approaches consistently outperforms the other. In particular, EPOSS leads to a slightly higher hypervolume value for Epigenomics and Montage, while Generic does slightly better with SIPHT, CyberShake and LIGO. Essentially, there are no noticeable differences between the hypervolume with the two algorithms. However, there is a huge difference in the algoritm execution times, since EPOSS requires an execution time 70\% to 97\% lower than  Genetic-50 (88 to 98\% lower than  Genetic-100). In conclusion, EPOSS  provides solutions as good as Genetic, but with a dramatic reduction of the execution time.

\begin{figure}[t]
    \centering
    \includegraphics[width=0.7\columnwidth]{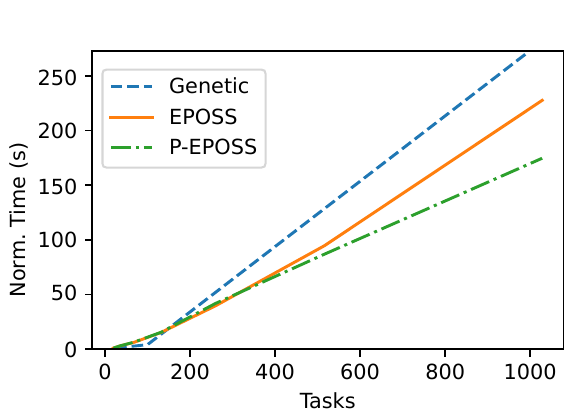}
     \vspace*{-0.3cm}\caption{Normalized algorithm execution times vs. workflow size with Genetic, EPOSS and P-EPOSS.}
    \label{fig:scalability1}
\end{figure}

\subsection{Scalability vs Workflow Size}

Finally, we show some data to compare the scalability of our algorithm and Genetic with respect to the workflow size. For Genetic, we used the scalability data provided by its authors in~\cite{calzarossa2021}, which refer to the Epigenomics workflow, whose size have been varied from 24 to 997 tasks. To compare the ability to scale of the two algorithms, we consider the execution time of each algorithm normalized to the execution time each of them required with the smallest size of the workflow (i.e. 24 tasks). We remark that our experiment showed that the execution times with Genetic are between 21 and 875 times higher than EPOSS, thus data we report in this section should be exclusively interpreted to the aim of comparing the capability to scale of each algorithm, rather than of comparing their performance. Figure~\ref{fig:scalability1} shows a comparison of the scalability curves achieved with Genetic, EPOSS and P-EPOSS. It appears clear that, starting from around 80 tasks, Genetic is subject to a more rapid increment with respect to EPOSS. Importantly, the parallel implementation of EPOSS appears to preserve linear scaling up to 1000 nodes. In fact, we observed that the incremental ratio between consecutive points of the curve of P-EPOSS stabilizes around the value of 1.61 from 150 tasks onward.

\section{Conclusions}
We proposed EPOSS, a workflow scheduling algorithm designed for IaaS cloud environments. It is based on a probabilistic 
formulation of the DAG scheduling problem, to cope with the 
performance uncertainty of cloud resources,
and includes constraints to model actual requirements of current IaaS providers. 
Therefore,
differently from most of the existing solutions, which are designed to operate with deterministic execution times, EPOSS can solve the more complex and computation-expensive probabilistic problem. Nevertheless, it keeps the overall computation cost largely affordable, and much lower than state-of-the-art probabilistic algorithms. We also presented a parallel version of EPOSS, which can further reduce the time required to compute a solution by taking advantage of hardware parallelism.

An extensive evaluation demonstrated that EPOSS is efficient and effective in a wide range of configuration scenarios, including different workflows, sets of VMs, task execution time distributions, scalability profiles, and various  constraints on the available resources. In particular, our results show that EPOSS can reduce 10 to over 100 times the time to find a solution compared to other probabilistic algorithms, with no loss in terms of solution quality. This represents a major goal for EPOSS, which ranks as the most efficient probabilistic algorithm currently in existence, hence being usable in a wide range of contexts.

As future work we plan to extend our study 
with automated strategies for profiling task execution times and distribution fitting, in order to design a framework able to fully automatize and optimize the workflow execution on IaaS clouds.

\ifCLASSOPTIONcompsoc
  \section*{Acknowledgments}
\else
  \section*{Acknowledgment}
\fi

The work presented in this article was partially funded by POR FESR Lazio 2014-2020 - "Progetti di Gruppi di Ricerca 2020" - project ID A0375-2020-36556.

\ifCLASSOPTIONcaptionsoff
  \newpage
\fi








\clearpage
\setcounter{figure}{0}
\setcounter{table}{0}
\setcounter{page}{1}
\section*{Supplemental Material} 
The material provided in the next pages is related to the article "Efficient Probabilistic Workflow Scheduling for IaaS Clouds" and includes the additional experimental data mentioned in Section \ref{sec:detailed_results} of the article.

Tables \ref{tab:mainCyberShake}--\ref{tab:mainSIPHT} show the detailed results achieved with the workflows CyberShake, LIGO, Montage and SIPHT, for $p_T$ equal to $0.75$, $0.9$ and $0.95$. We remark that column \textit{Hits} reports the percentage  of times the makespan was below the desired deadline for the solution found by the algorithm. If this percentage is less than or equal to $p_T$ then the solution $s$ is admissible -- i.e. the constraint $P(T(s) \leq d) < p_T$ of the problem formulation is met. In column \textit{Hits} the values that satisfy this requirement are highlighted in bold. Column \textit{Mon. Cost} shows the average monetary cost (in dollars) of the solutions found. Finally column {\em Exec. Time (s)} reports the algorithm execution time.

\setcounter{figure}{0}

\begin{table*}[]
\scriptsize
\centering
\caption{Results with the CyberShake workflow.}
\label{tab:mainCyberShake}
\begin{tabular}{rlrrr|rrr|rrr}
\toprule
 &  & \multicolumn{3}{c}{$p_T=0.75$} & \multicolumn{3}{c}{$p_T=0.9$} & \multicolumn{3}{c}{$p_T=0.95$} \\
VMs & Algorithm  & Hits (\%) & Mon. Cost (\$) &  Exec. Time(s) & Hits (\%) & Mon. Cost (\$) & Exec. Time (s) & Hits (\%) & Mon. Cost (\$) &  Exec. Time (s) \\
\midrule
\multirow[t]{9}{*}{$\Theta_{8}$} & GreedyCost & 0.0 & 0.23 & 0.03 & 0.0 & 0.23 & 0.03 & 0.0 & 0.23 & 0.03 \\
 & HEFT & \textbf{100.0} & 1.91 & 0.03 & \textbf{100.0} & 1.91 & 0.03 & \textbf{100.0} & 1.91 & 0.03 \\
 & MOHEFT & 8.7 & 0.57 & 3.15 & 8.7 & 0.57 & 3.13 & 8.7 & 0.57 & 3.17 \\
 & Dyna-50k & 0.0 & 0.94 & 3600.34 & 0.0 & 0.94 & 3600.32 & 0.0 & 0.94 & 3600.14 \\
 & Dyna-100k & 0.0 & 0.94 & 3600.19 & 0.0 & 0.94 & 3600.57 & 0.0 & 0.94 & 3600.58 \\
 & Genetic-50k & \textbf{78.5} & 1.38 & 1531.85 & \textbf{91.1} & 1.60 & 1541.00 & \textbf{95.1} & 1.58 & 1530.66 \\
 & Genetic-100k & 74.4 & 1.47 & 3223.00 & \textbf{100.0} & 2.28 & 3180.88 & 94.2 & 1.50 & 2986.10 \\
 & EPOSS & \textbf{79.5} & 0.52 & 27.08 & \textbf{96.2} & 0.98 & 28.15 & \textbf{99.3} & 0.86 & 26.89 \\
 & P-EPOSS & \textbf{97.5} & 0.76 & 19.63 & \textbf{98.9} & 0.98 & 19.65 & \textbf{97.1} & 0.69 & 19.49 \\
\midrule
\multirow[t]{7}{*}{$\Theta_{13}$} & GreedyCost & 0.0 & 0.23 & 0.05 & 0.0 & 0.23 & 0.05 & 0.0 & 0.23 & 0.05 \\
 & HEFT & \textbf{100.0} & 1.91 & 0.05 & \textbf{100.0} & 1.91 & 0.05 & \textbf{100.0} & 1.91 & 0.05 \\
 & MOHEFT & 3.3 & 1.91 & 10.94 & 3.3 & 1.91 & 10.96 & 3.3 & 1.91 & 10.91 \\
 & Genetic-50k & \textbf{87.3} & 2.11 & 1519.18 & 80.6 & 3.42 & 1506.21 & \textbf{97.0} & 2.02 & 1525.59 \\
 & Genetic-100k & \textbf{99.7} & 2.98 & 3097.73 & \textbf{94.8} & 1.91 & 3034.91 & \textbf{99.0} & 2.25 & 3047.05 \\
 & EPOSS & \textbf{90.8} & 0.61 & 47.28 & \textbf{93.5} & 1.46 & 45.29 & \textbf{99.3} & 1.11 & 46.74 \\
 & P-EPOSS & \textbf{98.4} & 0.76 & 35.18 & \textbf{98.4} & 0.76 & 35.24 & \textbf{98.4} & 0.76 & 35.34 \\
\midrule
\multirow[t]{7}{*}{$\Theta_{21}$} & GreedyCost & 0.0 & 0.23 & 0.10 & 0.0 & 0.23 & 0.10 & 0.0 & 0.23 & 0.10 \\
 & HEFT & \textbf{100.0} & 1.91 & 0.10 & \textbf{100.0} & 1.91 & 0.10 & \textbf{100.0} & 1.91 & 0.10 \\
 & MOHEFT & 16.3 & 0.55 & 21.22 & 16.3 & 0.55 & 21.37 & 16.3 & 0.55 & 21.35 \\
 & Genetic-50k & 73.2 & 1.58 & 1551.17 & 85.6 & 1.76 & 1544.36 & 81.6 & 3.23 & 1538.05 \\
 & Genetic-100k & \textbf{92.0} & 0.95 & 2996.63 & 66.0 & 6.71 & 3165.01 & \textbf{98.2} & 2.16 & 3018.90 \\
 & EPOSS & \textbf{97.3} & 0.86 & 91.41 & \textbf{97.3} & 0.86 & 91.70 & \textbf{98.6} & 0.77 & 91.63 \\
 & P-EPOSS & \textbf{96.8} & 0.60 & 76.08 & \textbf{96.8} & 0.60 & 76.28 & \textbf{96.8} & 0.60 & 75.66 \\
\bottomrule
\end{tabular}

\end{table*}

\begin{table*}[]
\scriptsize
\centering
\caption{Results with the LIGO workflow.}
\label{tab:mainLIGO}
\begin{tabular}{rlrrr|rrr|rrr}
\toprule
 &  & \multicolumn{3}{c}{$p_T=0.75$} & \multicolumn{3}{c}{$p_T=0.9$} & \multicolumn{3}{c}{$p_T=0.95$} \\
VMs & Algorithm  & Hits (\%) & Avg. Cost (\$) &  Exec. Time(s) & Hits (\%) & Avg. Cost (\$) & Exec. Time (s) & Hits (\%) & Avg. Cost (\$) &  Exec. Time (s) \\
\midrule
\multirow[t]{9}{*}{8} & GreedyCost & 0.0 & 1.16 & 0.03 & 0.0 & 1.16 & 0.03 & 0.0 & 1.16 & 0.03 \\
 & HEFT & \textbf{100.0} & 6.69 & 0.03 & \textbf{100.0} & 6.69 & 0.03 & \textbf{100.0} & 6.69 & 0.03 \\
 & MOHEFT & 63.2 & 1.76 & 5.17 & 63.2 & 1.76 & 5.13 & 63.2 & 1.76 & 5.18 \\
 & Dyna-50k & 0.0 & 1.80 & 3600.70 & 0.0 & 1.80 & 3600.15 & 0.0 & 1.80 & 3600.66 \\
 & Dyna-100k & 0.0 & 1.80 & 3600.41 & 0.0 & 1.80 & 3600.36 & 0.0 & 1.80 & 3600.11 \\
 & Genetic-50k & \textbf{77.0} & 2.69 & 2145.36 & \textbf{94.7} & 2.92 & 2124.81 & \textbf{97.8} & 5.99 & 2138.24 \\
 & Genetic-100k & \textbf{98.9} & 7.85 & 4517.37 & 87.5 & 2.69 & 4534.59 & \textbf{95.8} & 2.64 & 4648.29 \\
 & EPOSS & \textbf{90.9} & 2.04 & 57.94 & \textbf{99.5} & 2.24 & 57.67 & \textbf{99.5} & 2.24 & 57.39 \\
 & P-EPOSS & \textbf{92.2} & 2.20 & 27.21 & \textbf{93.5} & 2.03 & 27.30 & \textbf{99.5} & 1.89 & 27.31 \\
\midrule
\multirow[t]{7}{*}{13} & GreedyCost & 0.0 & 1.16 & 0.05 & 0.0 & 1.16 & 0.05 & 0.0 & 1.16 & 0.05 \\
 & HEFT & \textbf{100.0} & 6.69 & 0.05 & \textbf{100.0} & 6.69 & 0.05 & \textbf{100.0} & 6.69 & 0.05 \\
 & MOHEFT & 8.8 & 3.67 & 12.91 & 8.8 & 3.67 & 12.85 & 8.8 & 3.67 & 12.84 \\
 & Genetic-50k & \textbf{95.5} & 6.66 & 2116.25 & \textbf{90.6} & 3.11 & 2106.55 & \textbf{95.1} & 6.87 & 2080.25 \\
 & Genetic-100k & \textbf{93.0} & 2.79 & 4389.57 & 88.3 & 2.44 & 4412.43 & \textbf{99.6} & 5.01 & 4299.46 \\
 & EPOSS & \textbf{92.0} & 2.61 & 68.71 & \textbf{92.7} & 2.02 & 76.63 & \textbf{100.0} & 6.25 & 96.96 \\
 & P-EPOSS & \textbf{89.9} & 2.25 & 46.27 & \textbf{92.9} & 2.71 & 46.38 & \textbf{99.9} & 2.30 & 46.52 \\
\midrule
\multirow[t]{7}{*}{21} & GreedyCost & 0.0 & 1.16 & 0.09 & 0.0 & 1.16 & 0.09 & 0.0 & 1.16 & 0.09 \\
 & HEFT & \textbf{100.0} & 6.69 & 0.09 & \textbf{100.0} & 6.69 & 0.09 & \textbf{100.0} & 6.69 & 0.09 \\
 & MOHEFT & 4.1 & 6.90 & 25.28 & 4.1 & 6.90 & 25.42 & 4.1 & 6.90 & 25.28 \\
 & Genetic-50k & 72.4 & 3.12 & 2089.59 & \textbf{94.1} & 3.86 & 2029.39 & 94.7 & 3.15 & 2066.19 \\
 & Genetic-100k & \textbf{99.8} & 9.28 & 4369.91 & \textbf{99.8} & 5.77 & 4342.21 & \textbf{97.5} & 4.05 & 4337.04 \\
 & EPOSS & \textbf{84.1} & 2.34 & 141.77 & \textbf{94.4} & 2.25 & 146.98 & 94.4 & 2.25 & 146.93 \\
 & P-EPOSS & \textbf{90.2} & 1.96 & 105.20 & \textbf{95.5} & 2.36 & 104.96 & \textbf{100.0} & 2.18 & 104.41 \\
\bottomrule
\end{tabular}

\end{table*}

\begin{table*}[]
\scriptsize
\centering
\caption{Results with the Montage workflow.}
\label{tab:mainMontage}
\begin{tabular}{rlrrr|rrr|rrr}
\toprule
 &  & \multicolumn{3}{c}{$p_T=0.75$} & \multicolumn{3}{c}{$p_T=0.9$} & \multicolumn{3}{c}{$p_T=0.95$} \\
VMs & Algorithm  & Hits (\%) & Avg. Cost (\$) &  Exec. Time(s) & Hits (\%) & Avg. Cost (\$) & Exec. Time (s) & Hits (\%) & Avg. Cost (\$) &  Exec. Time (s) \\
\midrule
\multirow[t]{9}{*}{$\Theta_{8}$} & GreedyCost & 0.0 & 0.21 & 0.02 & 0.0 & 0.21 & 0.02 & 0.0 & 0.21 & 0.02 \\
 & HEFT & \textbf{99.9} & 1.19 & 0.02 & \textbf{99.9} & 1.19 & 0.02 & \textbf{99.9} & 1.19 & 0.02 \\
 & MOHEFT & 54.9 & 0.58 & 1.78 & 54.9 & 0.58 & 1.82 & 54.9 & 0.58 & 1.80 \\
 & Dyna-50k & 0.0 & 1.62 & 3600.24 & 0.0 & 1.62 & 3600.34 & 0.0 & 1.62 & 3600.65 \\
 & Dyna-100k & 0.0 & 1.62 & 3600.48 & 0.0 & 1.62 & 3600.33 & 0.0 & 1.62 & 3600.43 \\
 & Genetic-50k & 70.3 & 2.92 & 1248.41 & 86.2 & 3.51 & 1245.23 & 90.8 & 4.17 & 1259.34 \\
 & Genetic-100k & 67.4 & 1.87 & 2532.44 & \textbf{93.1} & 2.59 & 2468.63 & \textbf{95.2} & 3.01 & 2571.68 \\
 & EPOSS & \textbf{77.4} & 1.13 & 24.05 & \textbf{96.5} & 0.84 & 24.75 & \textbf{96.5} & 0.84 & 24.75 \\
 & P-EPOSS & \textbf{76.2} & 0.96 & 13.91 & \textbf{97.5} & 0.70 & 13.10 & \textbf{97.5} & 0.70 & 13.02 \\
\midrule
\multirow[t]{7}{*}{$\Theta_{13}$} & GreedyCost & 0.0 & 0.21 & 0.04 & 0.0 & 0.21 & 0.04 & 0.0 & 0.21 & 0.04 \\
 & HEFT & \textbf{99.9} & 1.19 & 0.04 & \textbf{99.9} & 1.19 & 0.04 & \textbf{99.9} & 1.19 & 0.04 \\
 & MOHEFT & 56.7 & 1.20 & 6.23 & 56.7 & 1.20 & 6.20 & 56.7 & 1.20 & 6.24 \\
 & Genetic-50k & \textbf{77.6} & 2.94 & 1201.08 & 87.7 & 3.54 & 1189.52 & 89.3 & 4.05 & 1182.60 \\
 & Genetic-100k & \textbf{87.1} & 2.65 & 2394.66 & \textbf{92.6} & 3.52 & 2467.76 & 84.6 & 5.05 & 2409.78 \\
 & EPOSS & \textbf{80.2} & 1.26 & 36.95 & \textbf{90.7} & 0.77 & 35.94 & \textbf{95.5} & 0.62 & 34.08 \\
 & P-EPOSS & \textbf{89.6} & 0.83 & 25.81 & \textbf{95.4} & 0.63 & 26.05 & \textbf{95.4} & 0.63 & 25.83 \\
\midrule
\multirow[t]{7}{*}{$\Theta_{21}$} & GreedyCost & 0.0 & 0.21 & 0.08 & 0.0 & 0.21 & 0.08 & 0.0 & 0.21 & 0.08 \\
 & HEFT & \textbf{99.9} & 1.19 & 0.07 & \textbf{99.9} & 1.19 & 0.07 & \textbf{99.9} & 1.19 & 0.07 \\
 & MOHEFT & 58.1 & 1.11 & 9.93 & 58.1 & 1.11 & 9.99 & 58.1 & 1.11 & 10.03 \\
 & Genetic-50k & 73.4 & 3.16 & 1234.89 & 86.4 & 4.06 & 1207.79 & 89.6 & 3.89 & 1207.80 \\
 & Genetic-100k & 71.0 & 2.55 & 2415.93 & 87.1 & 4.09 & 2373.94 & 92.6 & 2.38 & 2345.93 \\
 & EPOSS & \textbf{80.0} & 1.26 & 78.67 & \textbf{95.6} & 0.61 & 69.50 & \textbf{95.6} & 0.61 & 69.56 \\
 & P-EPOSS & \textbf{86.1} & 0.71 & 46.04 & \textbf{99.4} & 1.03 & 42.82 & \textbf{99.4} & 1.03 & 42.79 \\
\bottomrule
\end{tabular}

\end{table*}

\begin{table*}[]
\scriptsize
\centering
\caption{Results with the SIPHT workflow.}
\label{tab:mainSIPHT}
\begin{tabular}{rlrrr|rrr|rrr}
\toprule
 &  & \multicolumn{3}{c}{$p_T=0.75$} & \multicolumn{3}{c}{$p_T=0.9$} & \multicolumn{3}{c}{$p_T=0.95$} \\
VMs & Algorithm  & Hits (\%) & Avg. Cost (\$) &  Exec. Time(s) & Hits (\%) & Avg. Cost (\$) & Exec. Time (s) & Hits (\%) & Avg. Cost (\$) &  Exec. Time (s) \\
\midrule
\multirow[t]{9}{*}{$\Theta_{8}$} & GreedyCost & 0.8 & 0.61 & 0.02 & 0.8 & 0.61 & 0.02 & 0.8 & 0.61 & 0.02 \\
 & HEFT & \textbf{96.0} & 8.77 & 0.02 & \textbf{96.0} & 8.77 & 0.02 & \textbf{96.0} & 8.77 & 0.02 \\
 & MOHEFT & 49.8 & 1.05 & 2.23 & 49.8 & 1.05 & 2.30 & 49.8 & 1.05 & 2.25 \\
 & Dyna-50k & 1.9 & 4.02 & 3601.09 & 1.9 & 4.02 & 3601.59 & 1.9 & 4.02 & 3601.61 \\
 & Dyna-100k & 1.9 & 4.02 & 3601.53 & 1.9 & 4.02 & 3602.01 & 1.9 & 4.02 & 3601.26 \\
 & Genetic-50k & \textbf{84.0} & 1.68 & 3117.24 & \textbf{91.1} & 2.07 & 3083.85 & 94.6 & 2.83 & 3046.79 \\
 & Genetic-100k & \textbf{81.3} & 2.11 & 6500.74 & \textbf{92.5} & 3.11 & 6382.43 & \textbf{95.7} & 2.80 & 6308.65 \\
 & EPOSS & \textbf{76.6} & 0.96 & 21.29 & \textbf{91.8} & 1.66 & 21.37 & \textbf{96.0} & 9.24 & 20.62 \\
 & P-EPOSS & \textbf{76.6} & 0.96 & 16.57 & \textbf{93.1} & 1.17 & 16.12 & \textbf{95.3} & 1.38 & 15.98 \\
\midrule
\multirow[t]{7}{*}{$\Theta_{13}$} & GreedyCost & 0.8 & 0.61 & 0.04 & 0.8 & 0.61 & 0.04 & 0.8 & 0.61 & 0.04 \\
 & HEFT & \textbf{96.0} & 8.77 & 0.04 & \textbf{96.0} & 8.77 & 0.04 & \textbf{96.0} & 8.77 & 0.04 \\
 & MOHEFT & 49.8 & 1.05 & 7.39 & 49.8 & 1.05 & 7.34 & 49.8 & 1.05 & 7.37 \\
 & Genetic-50k & \textbf{82.1} & 2.21 & 3108.30 & \textbf{91.9} & 1.99 & 3083.70 & 94.8 & 4.01 & 3102.19 \\
 & Genetic-100k & \textbf{93.0} & 2.79 & 6470.66 & 87.6 & 1.50 & 6422.89 & 94.9 & 1.93 & 6289.47 \\
 & EPOSS & \textbf{76.6} & 0.96 & 31.23 & \textbf{91.8} & 1.66 & 33.25 & \textbf{96.0} & 9.02 & 30.85 \\
 & P-EPOSS & \textbf{76.6} & 0.96 & 25.37 & \textbf{93.1} & 1.17 & 23.41 & \textbf{95.5} & 1.37 & 23.33 \\
\midrule
\multirow[t]{7}{*}{$\Theta_{21}$} & GreedyCost & 0.8 & 0.61 & 0.07 & 0.8 & 0.61 & 0.07 & 0.8 & 0.61 & 0.07 \\
 & HEFT & \textbf{96.0} & 8.77 & 0.07 & \textbf{96.0} & 8.77 & 0.07 & \textbf{96.0} & 8.77 & 0.07 \\
 & MOHEFT & 49.8 & 1.05 & 14.18 & 49.8 & 1.05 & 14.12 & 49.8 & 1.05 & 14.14 \\
 & Genetic-50k & \textbf{86.6} & 2.55 & 3109.83 & \textbf{90.4} & 2.37 & 3117.49 & 94.2 & 2.50 & 3093.15 \\
 & Genetic-100k & \textbf{87.0} & 2.91 & 6348.05 & \textbf{95.9} & 3.36 & 6476.33 & \textbf{95.5} & 2.78 & 6309.11 \\
 & EPOSS & \textbf{76.6} & 0.96 & 72.72 & \textbf{91.8} & 1.66 & 78.42 & \textbf{96.3} & 11.03 & 70.41 \\
 & P-EPOSS & \textbf{76.6} & 0.96 & 62.90 & \textbf{93.1} & 1.16 & 64.96 & \textbf{95.5} & 1.37 & 64.36 \\
\bottomrule
\end{tabular}

\end{table*}

\begin{figure*}[h!]
    \centering
    \includegraphics[width=0.8\textwidth]{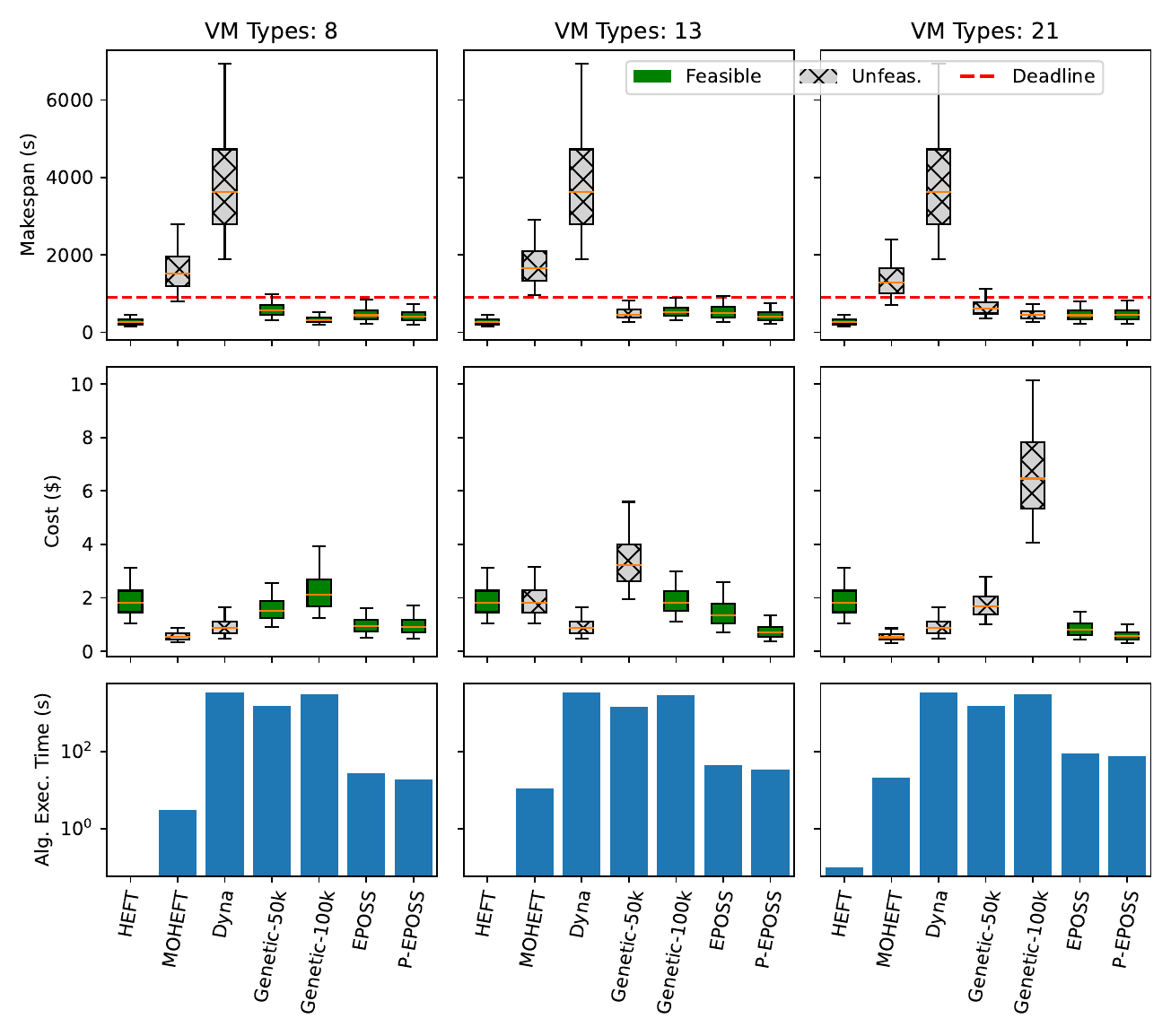}
    \caption{Graphical representation of the makespan, the monetary cost and the algorithm execution time with the CyberShake workflow ($p_T=0.9$)}
    \label{fig:mainCyberShake}
\end{figure*}

\begin{figure*}[h!]
    \centering
    \includegraphics[width=0.8\textwidth]{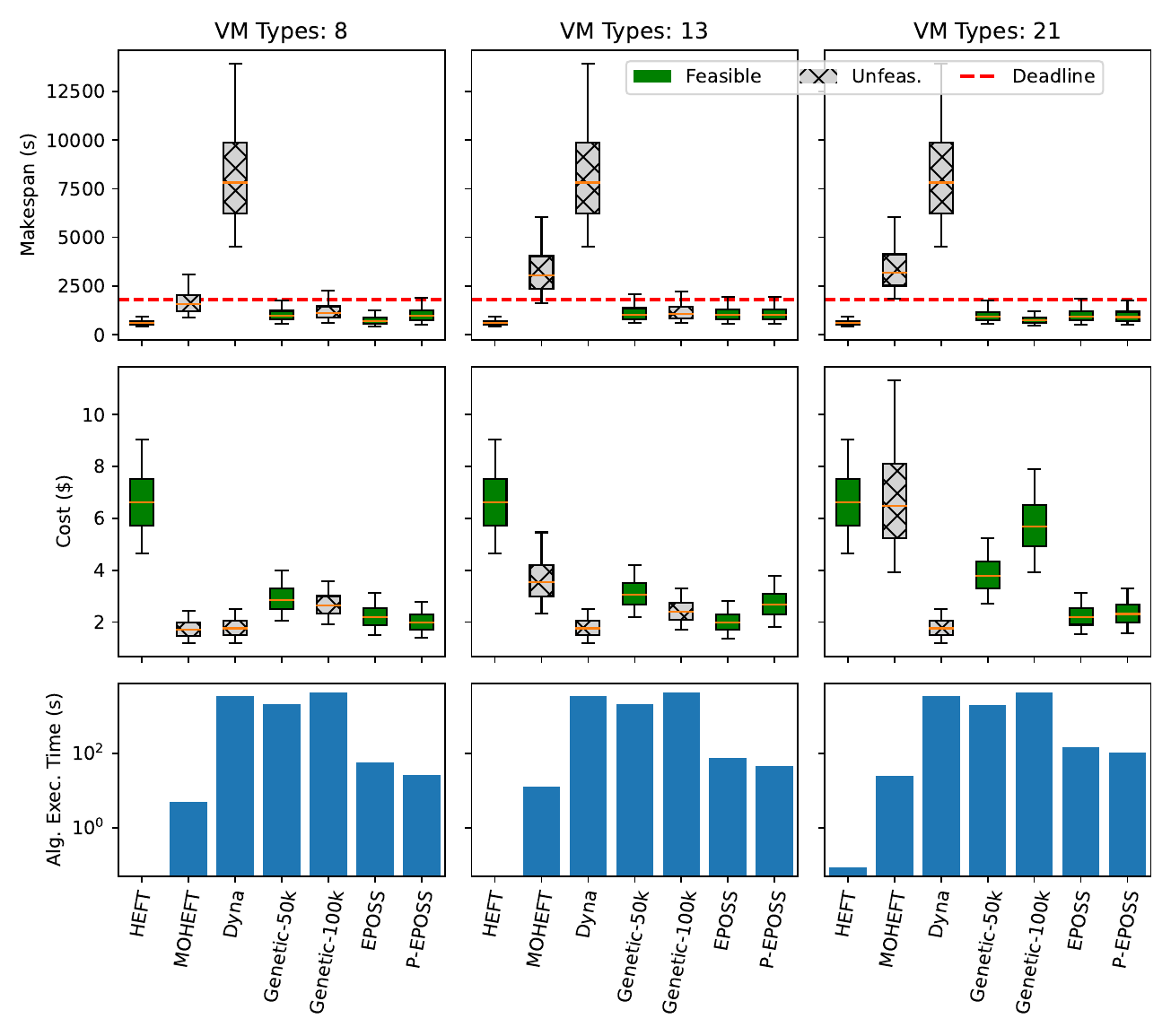}
    \caption{Graphical representation of the makespan, the monetary cost and the algorithm execution time with the LIGO workflow ($p_T=0.9$)}
        \label{fig:mainLIGO}
\end{figure*}

\begin{figure*}[h!]
    \centering
    \includegraphics[width=0.8\textwidth]{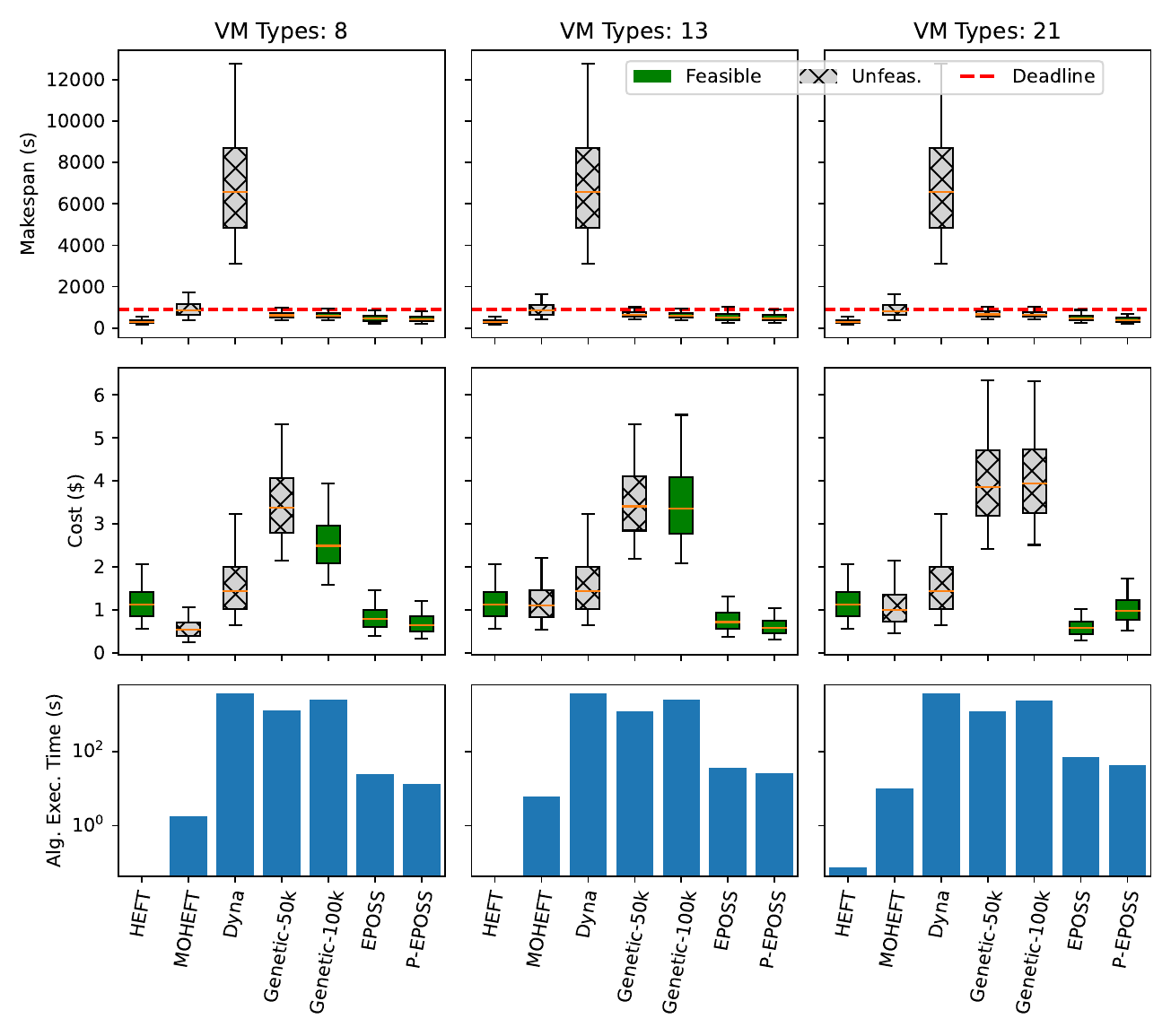}
    \caption{Graphical representation of the makespan, the monetary cost and the algorithm execution time with the Montage workflow ($p_T=0.9$)}
    \label{fig:mainMontage}
\end{figure*}

\begin{figure*}[h!]
    \centering
    \includegraphics[width=0.8\textwidth]{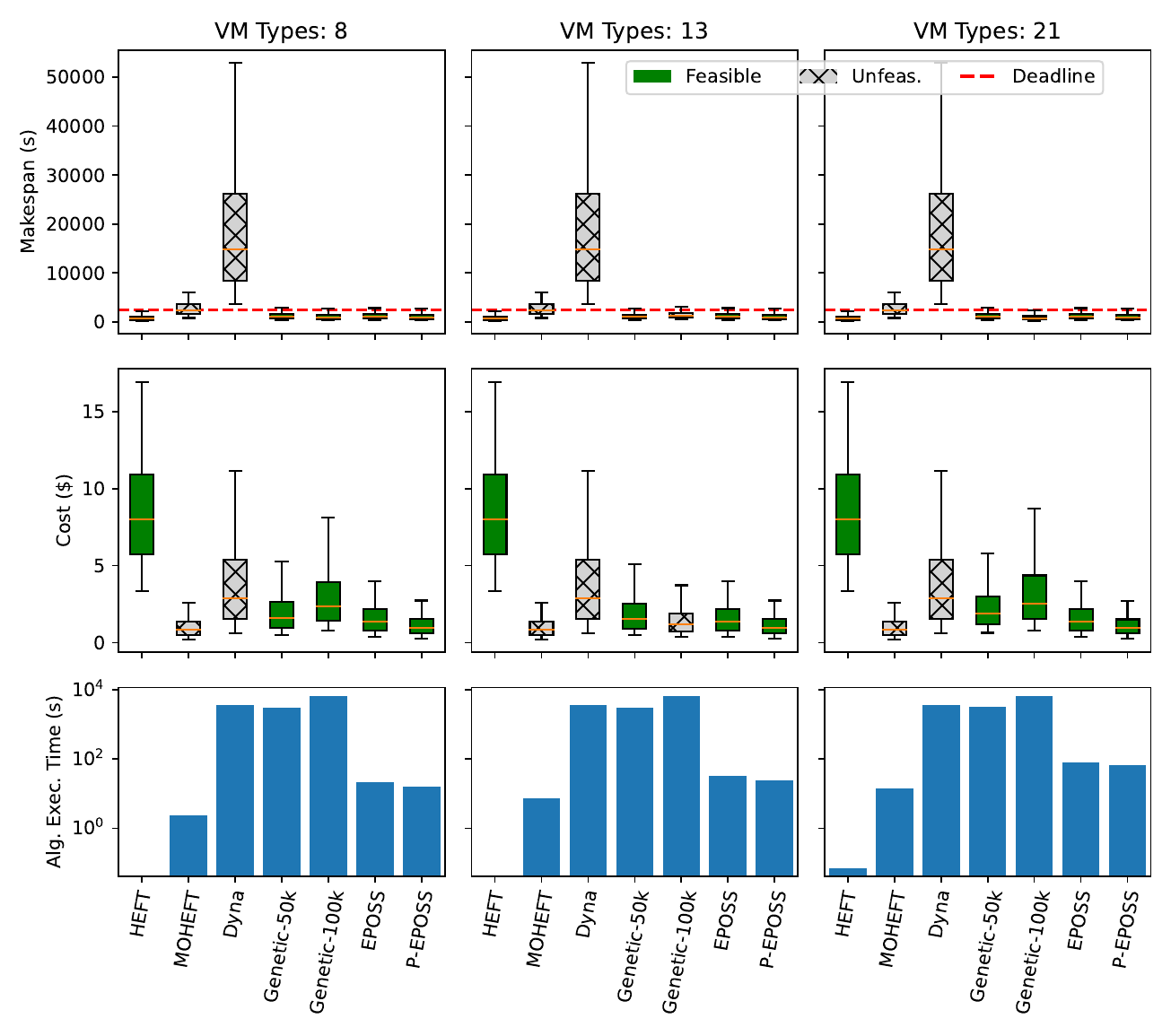}
    \caption{Graphical representation of the makespan, the monetary cost and the algorithm execution time with the SIPHT workflow ($p_T=0.9$)}
    \label{fig:mainSIPHT}    
\end{figure*}

Figures \ref{fig:mainCyberShake}--\ref{fig:mainSIPHT}
provide a graphical representation of 
makespan, monetary cost and algorithm execution time.
Specifically, for makespans and costs, the figures use boxes extending from the first to the third
quartile to depict value distribution, along with whiskers extending from
5$^{th}$ to 95$^{th}$ percentile and a horizontal line to indicate the median value.
Green boxes refer to scheduling solutions resulting in feasible executions (i.e. for which $P(T(s) \leq d)\geq0.9$), while gray bars refer to unfeasible solutions. The red dashed line marks the selected  deadline.


\remove{

\begin{figure*}[t]
    \centering
    \includegraphics[width=0.8\textwidth]{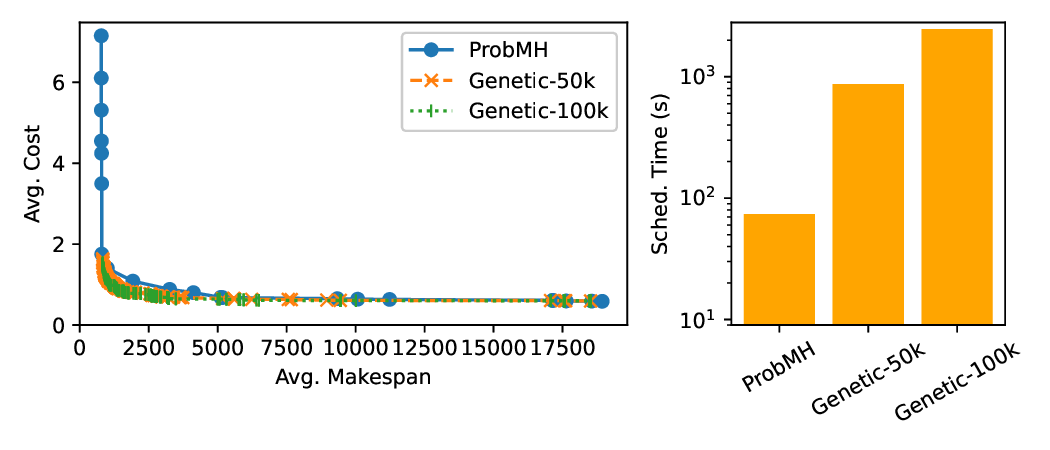}
    \caption{\hl{TODO} Front computation with SIPHT}
    \label{fig:frontierS}
\end{figure*}
\begin{figure*}[t]
    \centering
    \includegraphics[width=0.8\textwidth]{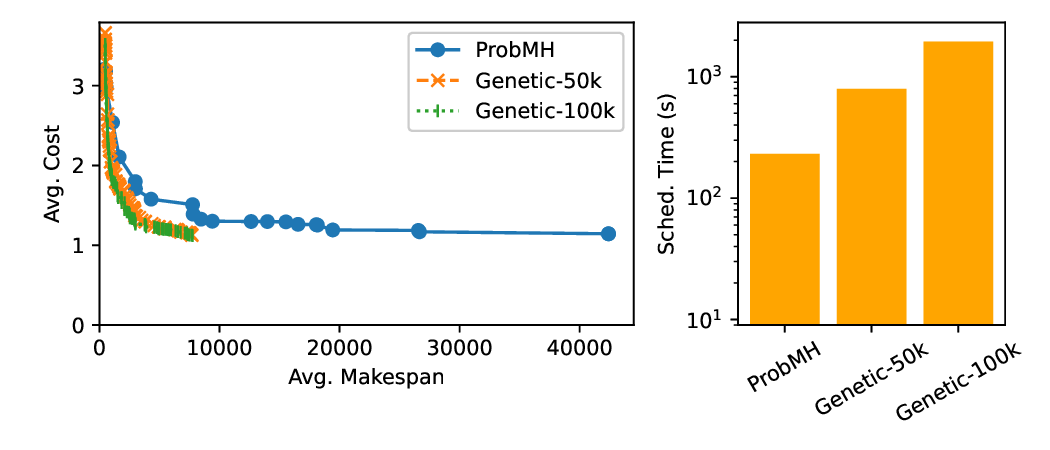}
    \caption{\hl{TODO} Front computation with LIGO}
    \label{fig:frontierL}
\end{figure*}
\begin{figure*}[t]
    \centering
    \includegraphics[width=0.8\textwidth]{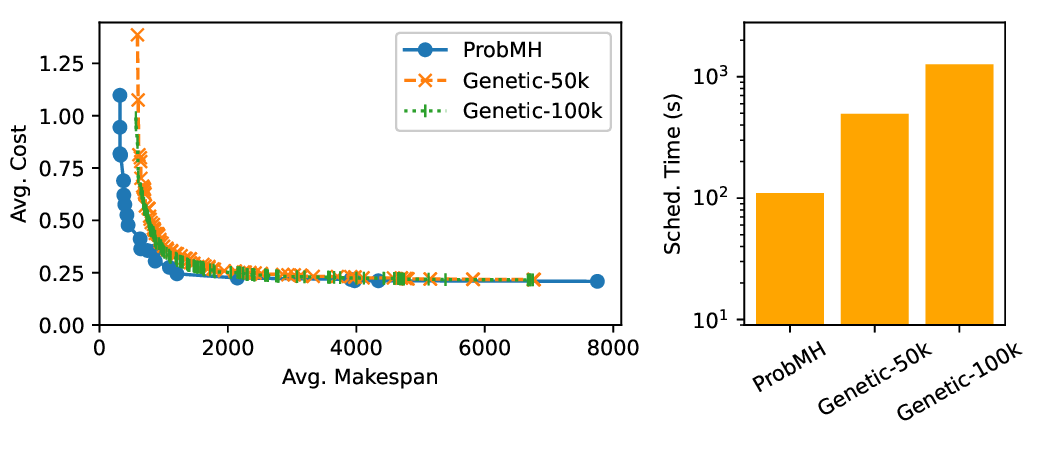}
    \caption{\hl{TODO} Front computation with Montage}
    \label{fig:frontierM}
\end{figure*}
\begin{figure*}[t]
    \centering
    \includegraphics[width=0.8\textwidth]{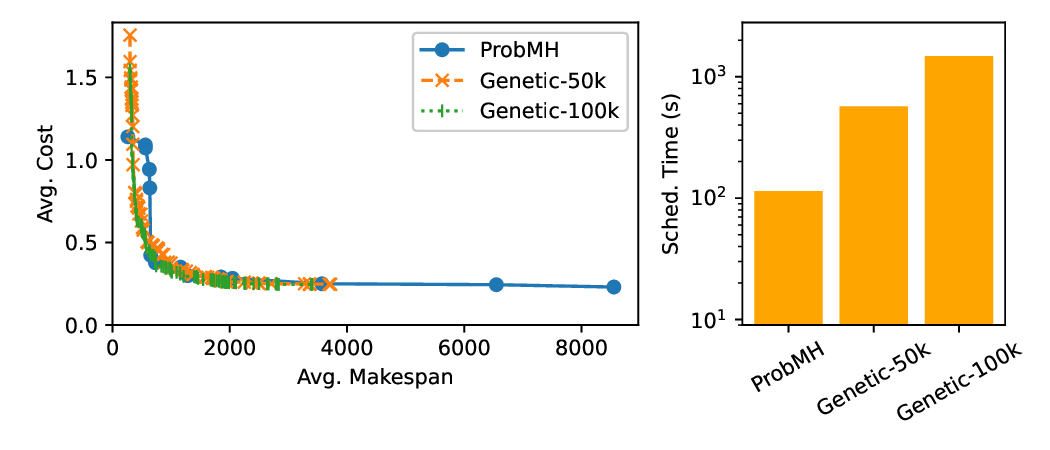}
    \caption{\hl{TODO} Front computation with CyberShake}
    \label{fig:frontierC}
\end{figure*}

}

\end{document}